\documentclass[a4paper,11pt]{revtex4}
\usepackage{graphicx} % standard LaTeX graphics tool;
\usepackage{amsmath} % For getting proper math eqns
\usepackage{amssymb} % Used for getting various symbols eg. gtrsim, lesssim etc.
\usepackage{bm} % For bold math (esp of lower greek letters)
\usepackage{dcolumn}% Align table columns on decimal point
\usepackage{color}
\usepackage{mathrsfs}
\usepackage{amsfonts}
\usepackage{varioref}
\usepackage{mathrsfs}
\usepackage{graphicx}
\usepackage{latexsym}
\usepackage{amsmath}
\usepackage{amssymb}
\usepackage{textcomp}
\usepackage{amsbsy}
\usepackage{graphics}
\usepackage{epstopdf}
\usepackage{color}
\usepackage[caption=false]{subfig}

\RequirePackage[colorlinks,citecolor=blue,urlcolor=magenta,linkcolor=blue]{hyperref}
\input epsf

\allowdisplaybreaks[4]
\newcommand{\e}{\mathrm{e}}

\begin{document}

\tolerance=5000

\title{Horizon entropy consistent with FLRW equations for general modified theories of gravity and for all EoS of the matter field}

\author{Shin'ichi~Nojiri$^{1,2}$\,\thanks{nojiri@gravity.phys.nagoya-u.ac.jp},
Sergei~D.~Odintsov$^{3,4}$\,\thanks{odintsov@ieec.uab.es},
Tanmoy~Paul$^{5,6}$\,\thanks{tpaul.phy@nitjsr.ac.in},
Soumitra~SenGupta$^{7}$\,\thanks{tpssg@iacs.res.in}} \affiliation{
$^{1)}$ Department of Physics, Nagoya University,
Nagoya 464-8602, Japan \\
$^{2)}$ Kobayashi-Maskawa Institute for the Origin of Particles
and the Universe, Nagoya University, Nagoya 464-8602, Japan \\
$^{3)}$ ICREA, Passeig Luis Companys, 23, 08010 Barcelona, Spain\\
$^{4)}$ Institute of Space Sciences (ICE, CSIC) C. Can Magrans s/n, 08193 Barcelona, Spain\\
$^{5)}$ National Institute of Technology Jamshedpur, Department of Physics, Jamshedpur - 831 014, India\\
$^{6)}$ Department of Physics, Visva-Bharati University, Santiniketan 731235\\
$^{7)}$ School of Physical Sciences, Indian Association for the Cultivation of Science, Kolkata-700032, India}

%\date{}

\tolerance=5000

\begin{abstract}
The question that continues to hinge the interrelation between cosmology and thermodynamics is broadly described as --- 
what is the form of horizon entropy that links the Friedmann equations for a ``$general$'' gravity theory with the underlying thermodynamics of the apparent horizon? 
The answer to this question was known only for Einstein's gravity and for $(n+1)$ dimensional Gauss-Bonnet gravity theory, 
but not for a general modified theory of gravity (for instance, the $F(R)$ gravity). 
In the present work, we take this issue and determine a general form of entropy that connects the Friedmann equations 
for any gravity theory with the apparent horizon thermodynamics given by $TdS = -dE + WdV$ (the symbols have their usual meaning in the context of entropic cosmology 
and $W = \left(\rho - p\right)/2$ is the work density of the matter fields represented by $\rho$ and $p$ as the energy density and the pressure, respectively). 
Using such generalized entropy, we find the respective entropies for several modified theories of gravity (including the $F(R)$ gravity). 
Further, it turns out that besides the above-mentioned question, the thermodynamic law $TdS = -dE + WdV$ itself has some serious difficulties for certain values of $\omega$ (the EoS of matter field). Thus we propose a modified thermodynamic law of apparent horizon, given by $TdS = -dE + \rho dV$, that is interestingly free from such difficulties. The modified law proves to be valid for all EoS of the matter field and thus is considered to be more general compared to the previous one which, however, is a limiting case of the modified law for $p = -\rho$.
Based on such modified thermodynamics, we further determine a generalized entropy that can provide the Friedmann equations of any general gravity theory 
for all values of EoS of the matter field. The further implications are discussed.
\end{abstract}
%%%%%%%%%%%%%%%%%%%%%%%%%%%%%%%%%%%%%%%%%%%%%%%%%%%%%%%%%%%%%%%%%%%%%%%%%%%%%%%%%%%%%%%%%%%%%%%%%%%
%%%%%%%%%%%%%%%%%%%%%%%%%%%%%%%%%%%%%%%%%%%%%%%%%%%%%%%%%%%%%%%%%%%%%%%%%%%%%%%%%%%%%%%%%%%%%%%%%%%
%%%%%%%%%%%%%%%%%%%%%%%%%%%%%%%%%%%%%%%%%%%%%%%%%%%%%%%%%%%%%%%%%%%%%%%%%%%%%%%%%%%%%%%%%%%%%%%%%%%
%\newpage
%%%%%%%%%%%%%%%%%%%%%%%%%%%%%%%%%%%%%%%%%%%%%%%%%%%%%%%%%%%%%%%%%%%%%%%%%%%%%%%%%%%%%%%%%%%%%%%%%%%
%%%%%%%%%%%%%%%%%%%%%%%%%%%%%%%%%%%%%%%%%%%%%%%%%%%%%%%%%%%%%%%%%%%%%%%%%%%%%%%%%%%%%%%%%%%%%%%%%%%
%%%%%%%%%%%%%%%%%%%%%%%%%%%%%%%%%%%%%%%%%%%%%%%%%%%%%%%%%%%%%%%%%%%%%%%%%%%%%%%%%%%%%%%%%%%%%%%%%%%
%\pacs{}

\maketitle

\section{Introduction}\label{SecI}

The discovery of Bekenstein-Hawking entropy associated with the event horizon of a black hole led us to think that gravity may have some connection with thermodynamics, 
otherwise, how does gravity know that the event horizon has an entropy and a temperature~\cite{Bekenstein:1973ur, Hawking:1975vcx}? 
Consequently, it was shown that the black hole mechanics derived from the gravitational field equations are equivalent to thermodynamic law(s) 
of the event horizon \cite{Bardeen:1973gs, Wald:1999vt, Jacobson:1995ab, Padmanabhan:2003gd, Padmanabhan:2009vy, Hayward:1997jp}. 
This interestingly brings two different arenas of physics, namely gravity, and thermodynamics, on an equal footing.

In the context of cosmology, the homogeneous and isotropic universe acquires an apparent horizon which, being a null surface, divides the observable universe from the unobservable one. 
Thus in the analogy of black hole thermodynamics, the apparent horizon in cosmology may also be associated with an entropy \cite{Cai:2005ra, Akbar:2006kj, Cai:2006rs, Akbar:2006er, 
Paranjape:2006ca, Sheykhi:2007zp, Jamil:2009eb, Cai:2009ph, Wang:2009zv, Jamil:2010di, Gim:2014nba, DAgostino:2019wko, Sanchez:2022xfh, Cognola:2005de, Brevik:2004sd}.
More explicitly, the cosmic entropy may be thought of as due to some energy flux of the matter fields (with the cosmic time) from inside to the outside of the horizon, 
and because of the null property of the horizon, such energy flux is regarded as some kind of information loss which in turn gives rise to an entropy of the apparent horizon. 
However, it deserves mentioning that a proper microscopic origin of the horizon entropy is still questionable (one may see \cite{Nojiri:2023ikl} for recent development in this regard). 
Initially, the apparent horizon is thought to have a Bekenstein-Hawking-like entropy (replacing the event horizon with the apparent horizon) which establishes the connection 
between the usual Friedmann equations with the underlying horizon thermodynamics given by $TdS = -dE + dW$ where $T$ is the temperature of the horizon, 
$E$ is the total internal energy of the matter fields inside of the horizon and $dW$ represents the work done by the matter fields \cite{Cai:2005ra, Akbar:2006kj}. 
However, the usual Friedmann equations with normal matter fields are unable to describe the dark energy era of the current universe. 
Thus, other kinds of entropies that are different from the Bekenstein-Hawking entropy have been proposed in the literature. 
Some of them are the Tsallis entropy \cite{Tsallis:1987eu}, the R{\'e}nyi entropy \cite{Renyi}, the Barrow entropy \cite{Barrow:2020tzx}, the Sharma-Mittal entropy \cite{SayahianJahromi:2018irq}, 
the Kaniadakis entropy \cite{Kaniadakis:2005zk, Drepanou:2021jiv}, the Loop Quantum Gravity entropy \cite{Majhi:2017zao, Liu:2021dvj} etc. 
Recently some generalized entropies (having 4 and 6 parameters respectively) have been proposed, which can generalize all such entropies known so far 
for suitable parameter spaces \cite{Nojiri:2022aof, Nojiri:2022dkr, Odintsov:2023qfj}. 
With the aid of horizon thermodynamics, the aforementioned entropies along with the generalized one introduce certain modification(s) to the usual Friedmann equations, 
which proves to have rich consequences in various aspects of cosmology, particularly in the dark energy sector \cite{Li:2004rb, Li:2011sd, Wang:2016och, Pavon:2005yx, Nojiri:2005pu, 
Nojiri:2020wmh, Gong:2004cb, Khurshudyan:2016gmb, Nojiri:2022nmu, Nojiri:2021jxf, Landim:2015hqa, Gao:2007ep, Li:2008zq, Nojiri:2023nop}.

Despite such advancements, the basic loophole that remains in entropic cosmology is to determine the correct horizon entropy that bridges the horizon thermodynamics 
with the Friedmann equations for a ``$general$'' gravity theory, and confirms the inter-relation between cosmology and thermodynamics. 
The answer to this was known for Einstein's gravity and $(n+1)$ dimensional Gauss-Bonnet theory, but not for a general modified theory of gravity. 
Actually, for Einstein's gravity as well as for $(n+1)$ dimensional Gauss-Bonnet theory, the cosmic entropy of the apparent horizon looks similar to that used 
in black hole thermodynamics for respective theories. 
However, the scenario becomes different for other modified gravity theories, for instance, the $F(R)$ gravity. 
In particular, the black hole-like entropy of $F(R)$ gravity is not able to produce the Friedmann equations of $F(R)$ theory from the cosmic thermodynamic law, 
and thus the correct horizon entropy for $F(R)$ cosmology remains unknown. 
In this regard, we would like to mention that the authors of \cite{Akbar:2006er} used a different thermodynamic law like 
$\left(\frac{1}{2\pi R_\mathrm{h}}\right)dS = -dE + dW$, and argued that the Bekenstein-Hawking entropy may be the correct horizon entropy even
for $F(R)$ cosmology, where $R_\mathrm{h}$ is the radius of the apparent horizon, and the suffix `eff' represents the effective energy-momentum tensor coming from 
both the matter field and the higher curvature terms as well. 
However, because the higher curvature contribution to the energy-momentum tensor is different for different modified gravity theories, 
the horizon thermodynamic law $\left(\frac{1}{2\pi R_\mathrm{h}}\right)dS = -dE + dW$ also gets changed with the gravitational theory.
This is not appropriate as the underlying thermodynamics of the apparent horizon should remain the same for any theory of gravity under consideration. 
Moreover, the factor $1/\left(2\pi R_\mathrm{h}\right)$ acts as the temperature of the apparent horizon in \cite{Akbar:2006er},
which is not valid for dynamical cases in the context of cosmology as also argued in \cite{Akbar:2006kj}. 

Therefore the natural question that remains in the context of entropic cosmology:
\begin{itemize}
\item What is the form of horizon entropy in cosmology, that links the underlying horizon thermodynamics with the Friedmann equations of a ``general'' gravity theory?
\end{itemize}
In the present work, we will address this issue by considering such thermodynamic law that remains the same in all gravity theories, and moreover, 
the horizon temperature will be proportional to the surface gravity of the dynamical apparent horizon. 
Before we proceed further, a few clarifications concerning the conventions and notations that we shall adopt are in order. 
Throughout this work we shall work in the geometric units with $G = 1$, $\hbar = 1$, $k_\mathrm{B}=1$ and $c = 1$. We will adopt the signature of the metric to be $(-,+,+,+)$. With this metric signature, we will assume the background to be the spatially flat Friedmann-Lema\^{i}tre-Robertson-Walker (FLRW) line element described
by the scale factor $a(t)$ and the Hubble parameter $H(t)$, with $t$ being the cosmic time. 
Moreover, $\rho$ and $p$ represent the homogeneous energy density and the homogeneous pressure respectively of the matter field(s) under consideration. 
Such matter field(s) may be contributed from some perfect fluid (in the case of standard Big-Bang cosmology) or even contributed from a scalar field (in the case of inflation). Hereafter we will use $\mathcal{Z}$ as the symbol for the action, while $S$ will be reserved to represent the horizon entropy.

\section{Thermodynamics of apparent horizon}\label{SecII}

We consider the $(n+1)$ dimensional spatially flat FLRW universe, whose metric is given by,
\begin{align}
\label{dS7}
ds^2 = \sum_{\mu,\nu=0,1,2,..,n} g_{\mu\nu} dx^\mu dx^\nu = - dt ^2 + a( t )^2 \left( d r ^2 + r ^2 {d\Omega_{n-1}}^2 \right) \, ,
\end{align}
where ${d\Omega_{n-1}}^2$ is the line element of an $(n-1)$ dimensional sphere of unit radius (particularly on the surface of the sphere). 
We also define
\begin{align}
\label{dS7B}
d{s_\perp}^2 = \sum_{M,N=0,1} h_{\mu\nu} dx^M dx^N = - dt ^2 + a( t )^2 d r ^2 \, .
\end{align}
The radius of the apparent horizon $R_\mathrm{h}=R\equiv a(t)r$ for the FLRW universe is given by the solution of the equation 
$h^{MN} \partial_M R \partial_N R = 0$ (see \cite{Cai:2005ra, Akbar:2006kj, Sanchez:2022xfh}) which immediately leads to, 
\begin{align}
\label{dS14A}
R_\mathrm{h}=\frac{1}{H}\, ,
\end{align}
with $H\equiv \frac{1}{a}\frac{da}{d t }$ represents the Hubble parameter of the universe. It may be noted that the apparent horizon in the case of a spatially flat FLRW universe 
becomes equal to the Hubble radius. 
The surface gravity $\kappa$ on the apparent horizon is defined as \cite{Cai:2005ra}
\begin{align}
\label{SG3}
\kappa= \left. \frac{1}{2\sqrt{-h}} \partial_M \left( \sqrt{-h} h^{MN} \partial_N R \right) \right|_{R=R_\mathrm{h}}\, .
\end{align}
For the metric of Eq.~(\ref{dS7}), we have $R=a r $ and obtain 
\begin{align}
\label{SG2}
\kappa = - \frac{1}{R_\mathrm{h}} \left\{ 1 + \dot{H}\left(\frac{{R_\mathrm{h}}^2}{2}\right) \right\} \, ,
\end{align}
where the following expression is used, 
\begin{align}
\label{dS14AB}
\dot R_\mathrm{h} = - H\dot{H} {R_\mathrm{h}}^3 \, ,
\end{align}
The surface gravity of Eq.~(\ref{SG2}) is related with the temperature via $T_\mathrm{h} = \kappa/(2\pi)$, i.e., 
\begin{align}
\label{AH2}
T_\mathrm{h} \equiv \frac{\left| \kappa \right|}{2\pi} 
= \frac{1}{2\pi R_\mathrm{h}} \left| 1 - \frac{\dot R_\mathrm{h}}{2 H R_\mathrm{h}} \right|
= \frac{H}{2\pi} \left| 1 + \frac{\dot{H}}{2H^2} \right|\, ,
\end{align}
in terms of the Hubble parameter and its derivative. 
Consequently, we may associate an entropy ($S$) to the apparent horizon, which in turn follows the thermodynamic law given by \cite{Akbar:2006kj, Sanchez:2022xfh},
\begin{align}
TdS = -dE + WdV\, ,
\label{law-1}
\end{align}
where $V = \Omega_nR_\mathrm{h}^n$ is the volume of the space enclosed by the horizon with $\Omega_n = \frac{\pi^{n/2}}{\Gamma\left[1+ n/2\right]}$ 
which represents the volume of a unit sphere. 
Moreover, $E = \rho V$ is the total internal energy of the matter fields inside of the horizon, and $W = \frac{1}{2}\left(\rho - p\right)$ denotes the work density 
by the matter fields \cite{Akbar:2006kj, Sanchez:2022xfh}. Eq.~(\ref{law-1}) clearly demonstrates that the horizon entropy generates due to --- 
(1) the decreasing of the total internal energy of the matter contents inside of the horizon, represented by the term $-dE$, and 
(2) the work done by the matter fields, given by $WdV$. 
Such decreasing internal energy and the work done may be effectively thought of as some energy flux of the matter fields from inside to the outside of the horizon 
for an infinitesimal time $dt$. Since the cosmic horizon, namely, $r_\mathrm{H} = \frac{1}{H}$, divides the observable universe from the unobservable one, 
the effective energy flux of the matter fields from inside to the outside of the horizon can be associated with some information loss which in turn gives rise to the entropy of the horizon.

\section{Consistent entropy for Einstein's gravity}\label{law-1-Einstein}

The $(n+1)$ dimensional FLRW equations in Einstein's gravity are (in the geometric units),
\begin{align}
H^2 = \frac{16\pi}{n(n-1)}~\rho\, ,\quad 
\dot{H} = -\frac{8\pi}{(n-1)}\left(\rho + p\right)\, .
\label{FRW-Einstein}
\end{align}
The matter fields (associated with $\rho$ and $p$) in the scenario obey the conservation law as,
\begin{align}
\dot{\rho} + nH\left(\rho + p\right) = 0\, ,
\label{conservation law}
\end{align}
To find the correct entropy that connects the above FLRW equations with the first law of thermodynamics, we start from Eq.~(\ref{law-1}) for the infinitesimal cosmic time $dt$, i.e.,
\begin{align}
T\frac{dS_\mathrm{E}}{dt} = -\frac{dE}{dt} + W\frac{dV}{dt}\, ,
\label{E-0}
\end{align}
This is a first-order differential equation with respect to $S_\mathrm{E}$ (where the suffix `E' is for Einstein's gravity), which we need to solve. 
For this purpose, we calculate the right-hand side of Eq.~(\ref{E-0}), as given by,
\begin{align}
 -\frac{dE}{dt} + W\frac{dV}{dt} = \left\{-\dot{\rho}V - \frac{1}{2}\left(\rho + p\right)\dot{V}\right\}\, ,
\label{E-1}
\end{align}
where we use $E = \rho V$. 
Due to the conservation Eq.~(\ref{conservation law}), the above equation turns out to be,
\begin{align}
 -\frac{dE}{dt} + W\frac{dV}{dt} = \left(\rho + p\right)\left(nHV - \dot{V}/2\right)
= 2\pi n\Omega_nR_\mathrm{h}^n\left(\rho + p\right)\left[\frac{H}{2\pi}\left(1 + \frac{\dot{H}}{2H^2}\right)\right]\, ,
\label{E-2}
\end{align}
where, in the second equality, we use $V = \Omega_nR_\mathrm{h}^n$. 
The term $(\rho+p)$ sitting in the above expression may be expressed in terms of $\dot{H}$ from the FLRW equation, or equivalently, 
in terms of $\dot{R}_\mathrm{h}$ by using $\dot{R}_\mathrm{h} = -\frac{\dot{H}}{H^2}$. 
Doing so, Eq.~(\ref{E-2}) leads to the following expression:
\begin{align}
 -\frac{dE}{dt} + W\frac{dV}{dt} = \frac{H}{2\pi}\left(1 + \frac{\dot{H}}{2H^2}\right)\frac{d}{dt}\left(\frac{n}{4}\Omega_n {R_\mathrm{h}}^{n-1}\right)\, ,
\label{E-3}
\end{align}
owing to which, Eq.~(\ref{E-0}) turns out to be,
\begin{align}
T\frac{dS_\mathrm{E}}{dt} = \frac{H}{2\pi}\left(1 + \frac{\dot{H}}{2H^2}\right)\frac{d}{dt}\left(\frac{n}{4}\Omega_n{R_\mathrm{h}}^{n-1}\right)\, ,
\label{E-4}
\end{align}
By using Eq.~(\ref{AH2}), we may solve the above equation to get $S_\mathrm{E}$ as,
\begin{align}
S_\mathrm{E} = \frac{n}{4}\Omega_n{R_\mathrm{h}}^{n-1} = \frac{A}{4}\, ,
\label{correct entropy-Einstein}
\end{align}
where $A = n\Omega_n{R_\mathrm{h}}^{n-1}$ represents the area of the apparent horizon.
Therefore the entropy, that establishes the connection between the FLRW equations of Einstein's gravity and the thermodynamic law (\ref{law-1}),
is given by $S_\mathrm{E} = \frac{A}{4}$. Such area law of the horizon entropy in Einstein's gravity is also proposed by the authors of \cite{Sanchez:2022xfh}, however in $(3+1)$ dimensional spacetime. The important point that Eq.~(\ref{correct entropy-Einstein}) shows is the horizon entropy corresponding to the Einstein's gravity still goes by $\frac{A}{4}$ even in $(n+1)$ dimensional spacetime, i.e., the form of horizon entropy for Einstein's gravity does not depend on the dimension of spacetime.

The form of $S_\mathrm{E}$ seems to be like the Bekenstein-Hawking entropy which was initially proposed for black hole thermodynamics, however,
the only difference from black hole thermodynamics is that the horizon of a black hole is static, unlike the cosmological scenario where the apparent horizon has a dynamical nature.

Here it deserves mentioning that to solve Eq.~(\ref{E-4}), we consider the term $\left(1 + \frac{\dot{H}}{2H^2}\right)$ to be positive 
(as the temperature contains its absolute value), otherwise, the entropy turns out to be negative which is impossible. In this regard one may think that by considering a positive integration constant upon the integration of Eq.~(\ref{E-4}), the entropy can be made positive valued even for $\left(1 + \frac{\dot{H}}{2H^2}\right) < 0$. However this will not be the case, as we now demonstrate: under the condition of $\left(1 + \frac{\dot{H}}{2H^2}\right) < 0$, the solution of Eq.~(\ref{E-4}) becomes, $S_\mathrm{E} = C - \frac{A}{4}$ where $C$ is an integration constant. To determine $C$, we may impose the condition that the $entropy~must~be~positive$, which is indeed a viable condition as entropy of a system is generally defined by the logarithmic of accessible microstates of the system.  Therefore in order to have a positive horizon entropy, the constant $C$ must take such value that $S_\mathrm{E} > 0$ for all possible $A$. Since the maximum of $A$ can go to infinity (when $H = 0$), the integration constant must take $C \rightarrow \infty$ which is not physical at all. Thus for the universe's evolution where $\left(1 + \frac{\dot{H}}{2H^2}\right) < 0$, there exists no such horizon entropy that connects the cosmology with the apparent horizon thermodynamics given by Eq.~(\ref{law-1}). The condition $1 + \frac{\dot{H}}{2H^2} < 0$ in turn leads to $\omega > \left(\frac{4}{n}-1\right)$ from the FLRW Eq.~(\ref{FRW-Einstein}),
where $\omega = \frac{p}{\rho}$ is the EoS parameter of the matter fields. 
In the case of $(3+1)$ dimensional spacetime, $\omega > \left(\frac{4}{n}-1\right)$ becomes $\omega > \frac{1}{3}$. In this regard, we may recall the reheating process in standard $(3+1)$ dimensional scalar field cosmology where the scalar field EoS during the reheating may become larger than $\frac{1}{3}$ for some suitable scalar field potential.
One may argue that for such reheating era where the $\omega$ becomes larger than $\frac{1}{3}$, or equivalently $1 + \frac{\dot{H}}{2H^2} < 0$, 
there exists no such entropy (of the horizon) that connects the FLRW Eq.~(\ref{FRW-Einstein}) with the thermodynamic law (\ref{law-1}).

\section{Consistent entropy for Gauss-Bonnet gravity}\label{sec-law-1-GB}

Let us start with the action for $(n+1)$ dimensional Gauss-Bonnet (GB) theory of gravity as,
\begin{align}
\mathcal{Z} = \int d^{n+1}x\sqrt{-g}\left[R + \widetilde{\alpha}\mathcal{G}\right] + \mathcal{Z}_\mathrm{mat}\, ,
\label{GB-action}
\end{align}
where $\mathcal{Z}_\mathrm{mat}$ is the action of matter field(s), $\mathcal{G} = R^2 - 4R_{\mu\nu}R^{\mu\nu} + R_{\mu\nu\alpha\beta}R^{\mu\nu\alpha\beta}$ and $\widetilde{\alpha}$
is the dimensionless GB parameter (recall that we are working in the unit of $8\pi G = 1$; and note that we use $\mathcal{Z}$ as the symbol for action, as $S$ is already reserved to represent the horizon entropy in the present context). For the metric ansatz in Eq.~(\ref{dS7}), the FLRW equations corresponding to the above action come as,
\begin{align}
H^2 + \alpha H^4 = \frac{16\pi}{n(n-1)}\rho\, ,\quad
\left(1 + 2\alpha H^2\right)\dot{H}= -\frac{8\pi}{(n-1)}\left(\rho + p\right)\, ,
\label{FRW-GB}
\end{align}
with $\alpha = (n-2)(n-3)\widetilde{\alpha}$. 
Therefore in $(3+1)$ dimensional spacetime, the parameter $\alpha$ vanishes and the above equations reduce to that of in Einstein's gravity --- this reflects the fact that 
in $(3+1)$ dimensional spacetime, the GB contribution becomes trivial and does not affect the equations of motion. 
Besides Eq.~(\ref{FRW-GB}), the matter field obeys the conservation law as stated in Eq.~(\ref{conservation law}). 
However all these three equations are not independent, actually one of them can be obtained from the other two.

The following entropy, in particular,
\begin{align}
S_\mathrm{GB} = \frac{A}{4}\left\{1 + \frac{2\alpha}{{R_\mathrm{h}}^2}\left(\frac{n-1}{n-3}\right)\right\}\, ,
\label{correct entropy-GB}
\end{align}
has been proposed in \cite{Akbar:2006kj} in order to link Eq.~(\ref{FRW-GB}) with the first law of thermodynamics. 
Here we are going to briefly review this, however, in a different way. 
It may be observed that $S_\mathrm{GB}$ gets a correction factor proportional to $R_\mathrm{h}^{n-3}$ (or $= 1/H^{n-3}$) over the Bekenstein-Hawking term. 
Thereby in $(3+1)$ dimensional spacetime, the correction term becomes a constant, and hence $S_\mathrm{GB}$ proves to be equivalent to the Bekenstein-Hawking entropy, 
in particular, $S_\mathrm{GB} \equiv \frac{A}{4}$ (note that $\frac{\alpha}{n-3} = (n-2)\widetilde{\alpha}$ which is not zero). 
This is a direct consequence of the fact that in $(3+1)$ dimensional spacetime, the FLRW equations of GB gravity reduce to that of Einstein's gravity 
where the corresponding entropy for the horizon is given by the Bekenstein-Hawking entropy.

The above form of entropy along with Eq.~(\ref{AH2}) immediately leads to the following expression of $TdS_\mathrm{GB}$ as,
\begin{align}
T\frac{dS_\mathrm{GB}}{dt} = -n\Omega_n{R_\mathrm{h}}^{n-1}\dot{H}\left(1 + 2\alpha H^2\right)\left|1 + \frac{\dot{H}}{2H^2}\right|\, ,\nonumber
\end{align}
which, due to Eq.~(\ref{FRW-GB}), can be rewritten as,
\begin{align}
T\frac{dS_\mathrm{GB}}{dt} = -n\Omega_n{R_\mathrm{h}}^{n-1}\rho\left(1 + \omega\right)\left|1 + \frac{\dot{H}}{2H^2}\right|\, ,
\label{GB-1}
\end{align}
where $\omega = p/\rho$ is the EoS of the matter field. 
Moreover, with the help of Eq.~(\ref{FRW-GB}), the change of total internal energy of the matter fields inside of the horizon and the amount of work done can be expressed by,
\begin{align}
\frac{dE}{dt}=&\, -n\Omega_n{R_\mathrm{h}}^{n-1}\rho\left[\left(1 + \omega\right) + \frac{\dot{H}}{H^2}\right]\, ,\nonumber\\
W\frac{dV}{dt}=&\, -n\Omega_n{R_\mathrm{h}}^{n-1}\rho\left(1 - \omega\right)\left(\frac{\dot{H}}{2H^2}\right)\, ,
\label{GB-2}
\end{align}
respectively, for an infinitesimal time $dt$. 
It is clearly evident that Eq.~(\ref{GB-1}) and Eq.~(\ref{GB-2}) result to the following thermodynamic relation between the extensive variables $(S,E,V)$ as,
\begin{align}
TdS_\mathrm{GB} = -dE + WdV
\label{GB-3}
\end{align}
provided $\left(1 + \frac{\dot{H}}{2H^2}\right) > 0$. 
Therefore based on the thermodynamic law (\ref{law-1}), the entropy $S_\mathrm{GB}$ proves to be the correct entropy that is consistent 
with the FLRW equations of $(n+1)$ dimensional GB gravity. Similar to the case of Einstein's gravity, an important condition to arrive 
at Eq.~(\ref{GB-3}) is $\left(1 + \frac{\dot{H}}{2H^2}\right) > 0$, otherwise $dS$ turns out to be negative, or equivalently, 
there exists no such entropy (of the horizon) that bridges the FLRW Eq.~(\ref{FRW-GB}) with the thermodynamic law.

\section{Consistent entropy for general modified theories of gravity}\label{sec-law-1-MGT}

The above two sections immediately lead to the following question: what is the form of entropy which, based on the thermodynamic law (\ref{law-1}), 
can produce the cosmological field equations for general modified theories of gravity? This is the subject of the present section.

A general modified theories of gravity (along with matter fields) in $(n+1)$ dimensional spacetime has the following action,
\begin{align}
\mathcal{Z} = \int d^{n+1}x~\sqrt{-g}~F \left( R,\mathcal{G},\cdots \right) + \mathcal{Z}_\mathrm{mat}\, ,
\label{MGT-action}
\end{align}
where $F \left(R,\mathcal{G}, \cdots \right)$ can be any analytic function of spacetime curvature and its higher order(s),
and $\mathcal{Z}_\mathrm{mat}$ represents the action for matter field(s) (with $\rho$ and $p$ are the associated energy density and the pressure respectively).
The gravitational equation for the above action is given by,
\begin{eqnarray}
 G_{\mu\nu} + Q_{\mu\nu} = 8\pi T_{\mu\nu}\, ,
 \label{new-1}
\end{eqnarray}
where $G_{\mu\nu}$ is the Einstein tensor (formed by the metric $g_{\mu\nu}$) and
\begin{eqnarray}
 T_{\mu\nu} = -\left(\frac{1}{\sqrt{-g}}\right)\frac{\delta\mathcal{Z}_\mathrm{mat}}{\delta g^{\mu\nu}}\, ,
 \label{new-2}
\end{eqnarray}
represents the energy-momentum tensor of the matter field(s). Moreover, $Q_{\mu\nu}$ is given by,
\begin{eqnarray}
 Q_{\mu\nu} = \left(\frac{1}{\sqrt{-g}}\right)\frac{\delta}{\delta g^{\mu\nu}}\left\{\sqrt{-g}\left(F \left( R,\mathcal{G}, \cdots \right) - R\right)\right\}\, .
 \label{new-3}
\end{eqnarray}
The term $Q_{\mu\nu}$ corresponds to the modification of gravitational action from the usual Einstein-Hilbert action. Here it deserves mentioning that $Q_{\mu\nu}$ is not treated as some energy-momentum tensor, rather it is driven solely by higher order(s) of spacetime curvature present in the gravitational action. The diffeomorphism invariance of the gravitational action (i.e. the Lie derivative of the gravitational action, generated by an arbitrary infinitesimal vector field, vanishes) leads to the Bianchi identity of the geometry \cite{Koivisto:2005yk,Boehmer:2021aji}, in particular, $\nabla_{\mu}\left(G^{\mu\nu} + Q^{\mu\nu}\right) = 0$ which in turn gives the conservation of energy-momentum tensor of the matter field(s) (i.e $\nabla_{\mu}T^{\mu\nu} = 0$) from Eq.~(\ref{new-1}). Regarding $\nabla_{\mu}\left(G^{\mu\nu} + Q^{\mu\nu}\right) = 0$ from the diffeomorphism invariance of a general modified gravity theory, one may see \cite{Koivisto:2005yk,Boehmer:2021aji}.

The FLRW equations coming from Eq.~(\ref{new-1}) can be expressed as,
\begin{align}
H^2=&\, \frac{16\pi}{n(n-1)}(\rho + \rho_\mathrm{c})\, ,\nonumber\\
\dot{H}=&\, -\frac{8\pi}{(n-1)}(\rho + \rho_\mathrm{c} + p + p_\mathrm{c})\, ,
\label{FRW-MGT}
\end{align}
Here $\rho_\mathrm{c}$ and $p_\mathrm{c}$ are the temporal component and the spatial component of $Q_{\mu\nu}$ respectively, i.e., $\rho_\mathrm{c}$ and $p_\mathrm{c}$ are not some energy-momentum tensor, rather these are curvature driven terms corresponding to the higher order curvature term(s) present in the gravitational action (\ref{MGT-action}). To get a better feeling about $\rho_\mathrm{c}$ and $p_\mathrm{c}$, we may consider the example of GB theory of gravity where
$F \left(R,\mathcal{G},\cdots \right) = R+\widetilde{\alpha}\mathcal{G}$ and the FLRW equations are given by Eq.~(\ref{FRW-GB}).
Comparison of Eq.~(\ref{FRW-GB}) with Eq.~(\ref{FRW-MGT}) leads to the $\rho_\mathrm{c}$ and $p_\mathrm{c}$ corresponding to the Gauss-Bonnet theory as follows,
%Here $\rho_\mathrm{eff} = \rho + \rho_\mathrm{c}$ (and $p_\mathrm{eff} = p+p_\mathrm{c}$) with $\rho_\mathrm{c}$ and $p_\mathrm{c}$
%denote the energy density and pressure coming from the higher curvature terms present in the gravitational action. In particular, $\rho_\mathrm{c}$ and $p_\mathrm{c}$ %correspond to the energy-momentum tensor given by,
%\begin{eqnarray}
% T_\mathrm{\mu\nu} = \left(\frac{2}{\sqrt{-g}}\right)\frac{\delta}{\delta g^{\mu\nu}}\left\{\sqrt{-g}\left(F \left( R,\mathcal{G},R_{\mu\nu}R^{\mu\nu},\cdots \right) - R\right)\right\}~~,\nonumber
%\end{eqnarray}
%clearly $T_\mathrm{\mu\nu}$ vanishes for Einstein gravity. Moreover the effective energy density ($\rho_\mathrm{eff}$) obeys the conservation law as,
%\begin{align}
%\dot{\rho}_\mathrm{eff} + nH\left(\rho_\mathrm{eff} + p_\mathrm{eff}\right) = 0\, ,
%\label{conservation law-MGT}
%\end{align}
\begin{align}
\rho_\mathrm{c} = -\left(\frac{n(n-1)}{16\pi}\right)\alpha H^4\, ,\quad
\rho_\mathrm{c} + p_\mathrm{c} = -\frac{2\alpha H^2\left(\rho + p\right)}{1+ 2\alpha H^2}\, ,
\label{ep-GB}
\end{align}
Clearly both the above $\rho_\mathrm{c}$ and $p_\mathrm{c}$ are proportional to the GB parameter $\alpha$, and 
thus they generate solely due to the presence of $\mathcal{G}$ in the gravitational action. 
Moreover, in the case of $(3+1)$ dimensional $F(R)$ gravity where the action is $\mathcal{Z} = \int d^4x \sqrt{-g}\left[F(R)\right]+ \mathcal{Z}_\mathrm{mat}$, i.e., $F\left(R,\mathcal{G},\cdots \right) = F(R)$,
the FLRW equations are given by,
\begin{align}
H^2=&\, \frac{8\pi}{3}\left\{\rho - \frac{f(R)}{2} + 3f'(R)\left(H^2 + \dot{H}\right) - 18f''(R)\left(4H^2\dot{H} + H\ddot{H}\right)\right\}\, ,\nonumber\\
\dot{H}=&\, -8\pi\left\{\rho + p + 2\dot{H}f'(R) + 6f''(R)\left(-4H^2\dot{H} + 4\dot{H}^2 + 3H\ddot{H} + \dddot{H}\right) + 36f'''(R)\left(4H\dot{H} + \ddot{H}\right)^2\right\}\, ,
\label{FRW-FR}
\end{align}
with $f(R) = (F(R) - R)$ and $f'(R) = \frac{df}{dR}$. 
By comparing the above set of equations with Eq.~(\ref{FRW-MGT}), we get
\begin{align}
\rho_\mathrm{c}=&\, - \frac{f(R)}{2} + 3f'(R)\left(H^2 + \dot{H}\right) - 18f''(R)\left(4H^2\dot{H} + H\ddot{H}\right)\, ,\nonumber\\
\rho_\mathrm{c} + p_\mathrm{c}=&\, 2\dot{H}f'(R) + 6f''(R)\left(-4H^2\dot{H} + 4\dot{H}^2 + 3H\ddot{H} + \dddot{H}\right) + 36f'''(R)\left(4H\dot{H} + \ddot{H}\right)^2\, ,
\label{ep-FR}
\end{align}
corresponding to the $F(R)$ gravity theory. 
Once again, $\rho_\mathrm{c} = p_\mathrm{c} = 0$ for $F(R) = R$ (i.e., for $f(R) = 0$) --- this indicates that the $\rho_\mathrm{c}$ and $p_\mathrm{c}$ 
arise solely due to the presence of higher curvature term(s) in the action.

Coming back to the general modified gravity theory described by the action (\ref{MGT-action}), the FLRW equations for general modified gravity theory get modified compared to the usual FLRW equations and the modifications are captured within $\rho_\mathrm{c}$ and $p_\mathrm{c}$ (see Eq.~(\ref{FRW-MGT})) which, as mentioned earlier, are the temporal and the spatial components of $Q_{\mu\nu}$ respectively. Here we want to find the horizon entropy that links the FLRW equations for a general modified gravity theory with the thermodynamic law of the apparent horizon. For this purpose, we may take some guess from the previous Sec.~[\ref{sec-law-1-GB}] (in the case of GB gravity theory) where we showed that the form of the horizon entropy deviates from $\frac{A}{4}$ due to the GB term in the gravitational action, in particular, the horizon entropy corresponding to the GB theory proves to be $\equiv \frac{A}{4} + \textrm{``a non-trivial term depending on the area (or the radius) of the apparent horizon''}$. Therefore for general modified gravity theory, we may expect that the required horizon entropy is given by,
\begin{align}
S_\mathrm{g} = \frac{A}{4} + S_\mathrm{c}(A)\, ,
\label{correct entropy-MGT-1}
\end{align}
with the suffix 'g' is for 'general' gravity theory. Here $A = n\Omega_n{R_\mathrm{h}}^{n-1}$ represents the area of the apparent horizon in $n+1$ dimensional spacetime, and $S_\mathrm{c}$ is a function of $A$. Regarding the above Eq.~(\ref{correct entropy-MGT-1}), we would like to mention that $S_\mathrm{g}$ does not separate the entropy from Einstein gravity and that from modified gravity of higher curvature terms. Actually both the terms in the R.H.S of Eq.~(\ref{correct entropy-MGT-1}) contain the information of the modification of gravitational action (implicitly or explicitly). Regarding the first term in the R.H.S of Eq.~(\ref{correct entropy-MGT-1}), the area of the apparent horizon goes by $1/H^2$, and the evolution of $H$ is indeed affected by $\rho_\mathrm{c}$ and $p_\mathrm{c}$ from Eq.~(\ref{FRW-MGT}) --- thus the term $\frac{A}{4}$ of Eq.~(\ref{correct entropy-MGT-1}) \textrm{implicitly} encapsulates the higher order curvature terms' information. On other side, $S_\mathrm{c}$ \textrm{explicitly} depends on the modification of gravitational action, which we need to find in such a way that the following thermodynamic law, namely
%Coming back to the general modified gravity theory described by the action (\ref{MGT-action}),
%the presence of higher-order curvature terms introduces a correction to the horizon entropy over the usual Bekenstein-Hawking one.
%Hence the entropy, that links the FLRW equations of a general modified gravity theory with the thermodynamic law, can be written as,
%\begin{align}
%S_\mathrm{g} = S_\mathrm{E} + S_\mathrm{c} = \frac{A}{4} + S_\mathrm{c}\, ,
%\label{correct entropy-MGT-1}
%\end{align}
%with the suffix 'g' is for 'general' gravity theory, and $S_\mathrm{E} = \frac{A}{4} = \frac{n}{4}\Omega_n{R_\mathrm{h}}^{n-1}$
%is the corresponding entropy for Einstein's gravity, see Eq.~(\ref{correct entropy-Einstein}).
\begin{align}
TdS_\mathrm{g} = -dE + WdV\, ,
\label{MGT-1}
\end{align}
holds, with $E = \rho V$ and $W = \frac{1}{2}\left(\rho - p\right)$.
By using $S_\mathrm{g} = \frac{A}{4} + S_\mathrm{c}$, the above equation can be equivalently written as,
\begin{align}
T\frac{dS_\mathrm{c}}{dt} = \frac{d}{dt}\left(\rho_\mathrm{c}V\right) - W_\mathrm{c}\frac{dV}{dt} 
+ \left[-\frac{d}{dt}\left(\rho V + \rho_\mathrm{c}V\right) + \frac{1}{2}\left(\rho + \rho_\mathrm{c} - p - p_\mathrm{c}\right)\frac{dV}{dt} - T\frac{d}{dt}\left(\frac{A}{4}\right)\right]\, ,
\label{MGT-2}
\end{align}
where $W_\mathrm{c} = \frac{1}{2}\left(\rho_\mathrm{c} - p_\mathrm{c}\right)$. By using $\nabla_{\mu}Q^{\mu\nu} = 0$ (as described after Eq.~(\ref{new-3})) along with $R_\mathrm{h} = \frac{1}{H}$, we determine the first two terms within the square bracket of the above expression as follows:
\begin{align}
 -\frac{d}{dt}\left(\rho V + \rho_\mathrm{c}V\right) + \frac{1}{2}\left(\rho + \rho_\mathrm{c} - p - p_\mathrm{c}\right)\frac{dV}{dt} = 2\pi n\Omega_nR_\mathrm{h}^n\left(\rho + \rho_\mathrm{c} + p + p_\mathrm{c}\right)\left[\frac{H}{2\pi}\left(1 + \frac{\dot{H}}{2H^2}\right)\right]\, ,
\label{MGT-3}
\end{align}
which, due to the FLRW Eq.~(\ref{FRW-MGT}) along with a bit of simplification, takes the following form,
\begin{align}
 -\frac{d}{dt}\left(\rho V + \rho_\mathrm{c}V\right) + \frac{1}{2}\left(\rho + \rho_\mathrm{c} - p - p_\mathrm{c}\right)\frac{dV}{dt} = \frac{H}{2\pi}\left(1 + \frac{\dot{H}}{2H^2}\right)\frac{d}{dt}\left(\frac{A}{4}\right)\, ,
\label{MGT-4}
\end{align}
By identifying $\frac{H}{2\pi}\left(1 + \frac{\dot{H}}{2H^2}\right) = T$ $\left(\mbox{provided}\ \left(1 + \frac{\dot{H}}{2H^2}\right) > 0 \right)$, namely the temperature of the horizon, 
we may note that the terms within the square bracket on the right-hand side of Eq.~(\ref{MGT-2}) cancel each other. 
As a result, Eq.~(\ref{MGT-2}) leads to,
\begin{align}
T\frac{dS_\mathrm{c}}{dt} = \frac{d}{dt}\left(\rho_\mathrm{c}V\right) - W_\mathrm{c}\frac{dV}{dt}\, .
\label{MGT-5}
\end{align}
For the case of the FLRW metric, the extended Bianchi identity $\nabla_{\mu}Q^{\mu\nu} = 0$ gives $\dot{\rho}_\mathrm{c} + nH\left(\rho_\mathrm{c} + p_\mathrm{c}\right) = 0$ (recall that $\rho_\mathrm{c}$and $p_\mathrm{c}$ are the temporal and the spatial components of $Q_{\mu\nu}$ respectively). Owing to this expression, Eq.~(\ref{MGT-5}) takes the following form,
\begin{align}
\frac{dS_\mathrm{c}}{dt} = -2\pi n\Omega_n\left(\rho_\mathrm{c} + p_\mathrm{c}\right){R_\mathrm{h}}^n\, ,
\label{MGT-6}
\end{align}
on integrating which, we get,
\begin{align}
S_\mathrm{c} = 2\pi n\Omega_n\int {R_\mathrm{h}}^{n-2}\left(\frac{\rho_\mathrm{c} + p_\mathrm{c}}{\dot{H}}\right)dR_\mathrm{h}\, ,
\label{MGT-7}
\end{align}
To arrive at Eq.~(\ref{MGT-7}), we use $\dot{R}_\mathrm{h} = -\frac{\dot{H}}{H^2}$. 
With the above form of $S_\mathrm{c}$, Eq.~(\ref{correct entropy-MGT-1}) provides the full entropy corresponding to a general modified gravity theory as,
\begin{align}
S_\mathrm{g} = \frac{A}{4} + 2\pi n\Omega_n\int {R_\mathrm{h}}^{n-2}\left(\frac{\rho_\mathrm{c} + p_\mathrm{c}}{\dot{H}}\right)dR_\mathrm{h}\, ,
\label{MGT-8}
\end{align}
It may be observed that for $\rho_\mathrm{c} = p_\mathrm{c} = 0$, i.e., for Einstein's gravity theory, the entropy from Eq.~(\ref{MGT-8}) becomes $S_\mathrm{g} = A/4$, as per our expectation.
The integration in the above expression cannot be realized without specifying the scale factor $a=a(t)$ as a function of the cosmological time
$t$ because $S_\mathrm{g}$ depends on $R_\mathrm{h}$ but not on $\dot H$.
Thus if we consider a specific scale factor $a(t)$, the Hubble parameter and consequently $R_\mathrm{h}$ is also given by a function of $t$ via Eq.~(\ref{dS14A}), 
$R_\mathrm{h}=R_\mathrm{h}(t)=\frac{1}{H(t)}$, which can be solved with respect
to $t$ as a function of $R_\mathrm{h}$, i.e., $t=t\left( R_\mathrm{h} \right)$. As a result, $\rho_\mathrm{c}$, $p_\mathrm{c}$ and $\dot H$ can be obtained as a function 
of $R_\mathrm{h}$, i.e., $\rho_\mathrm{c} = \rho_\mathrm{c}(R_\mathrm{h})$ and $\dot{H} = \dot{H}(R_\mathrm{h})$. 
Owing to this, Eq.~(\ref{MGT-8}) can be integrated to give $S_\mathrm{g}$.

Thus as a whole, Eq.~(\ref{MGT-8}) gives the correct horizon entropy ($S_\mathrm{g}$) that connects the FLRW equations of any gravity theories with the thermodynamic law (\ref{law-1}).
The form of $S_\mathrm{g}$ explicitly depends on the factor $\left(\frac{\rho_\mathrm{c} + p_\mathrm{c}}{\dot{H}}\right)$ which actually encodes the information 
of specific gravity theory under consideration. 
Below we will present some specific examples of gravity theories and will determine the respective entropy from Eq.~(\ref{MGT-8}).

\subsection{Case-I: In absence of matter fields}\label{sec-case-I}

For $\rho = p = 0$, i.e., without any matter fields, Eq.~(\ref{FRW-MGT}) gives,
\begin{align}
\frac{\rho_\mathrm{c} + p_\mathrm{c}}{\dot{H}} = -\frac{8\pi}{n-1}\, ,
\label{I-1}
\end{align}
Plugging the above expression into Eq.~(\ref{MGT-8}) immediately yields to
\begin{align}
S_\mathrm{g} = \frac{A}{4} - \frac{n(n-1)}{4}\Omega_n\int {R_\mathrm{h}}^{n-2}dR_\mathrm{h}\, ,
\label{I-2}
\end{align}
Due to $A = n\Omega_n{R_\mathrm{h}}^{n-1}$, the integral term in Eq.~(\ref{I-2}) exactly cancels with $\frac{A}{4}$, 
and thus the entropy becomes a constant appearing from the integration constant, i.e., $S_\mathrm{g} = \mathrm{constant}$ in this case. 
Therefore the entropy corresponding to any modified gravity theories {\it without} matter fields turns out to be a constant. 
This is, however, expected as there is no flux of matter fields from inside to outside of the horizon for $\rho = p = 0$, or equivalently 
there is no information loss associated with the horizon.

\subsection{Case-II: $(n+1)$ dimensional Gauss-Bonnet gravity}

In this subsection, we will consider $(n+1)$ dimensional Gauss-Bonnet gravity as an example of modified gravity theory, and will examine 
whether the generalized version of entropy found in Eq.~(\ref{MGT-8}) reduces to the form of $S_\mathrm{GB}$ (see Eq.~(\ref{correct entropy-GB})).

For $(n+1)$ dimensional GB gravity having the action (\ref{GB-action}), i.e., where $F \left( R,\mathcal{G},\cdots \right) = R + \widetilde{\alpha}\mathcal{G}$, 
the FLRW equations follow Eq.~(\ref{FRW-GB}) and the corresponding $\rho_\mathrm{c}$, $p_\mathrm{c}$ are obtained in Eq.~(\ref{ep-GB}). 
Therefore by using Eq.~(\ref{ep-GB}), we determine $\left(\rho_\mathrm{c} + p_\mathrm{c}\right)/\dot{H}$ in the case of GB gravity, and is given by,
\begin{align}
\frac{\rho_\mathrm{c} + p_\mathrm{c}}{\dot{H}} = -\frac{2\alpha H^2\left(\rho + p\right)}{\dot{H}\left(1+ 2\alpha H^2\right)}\, ,
\label{II-1}
\end{align}
which, due to the FLRW Eq.~(\ref{FRW-GB}), turns out to be,
\begin{align}
\frac{\rho_\mathrm{c} + p_\mathrm{c}}{\dot{H}} = \frac{\alpha(n-1)}{4\pi}H^2\, ,
\label{II-2}
\end{align}
Therefore in the case of GB gravity, the factor $\left(\rho_\mathrm{c} + p_\mathrm{c}\right)/\dot{H}$ does not depend on the derivative(s) of the Hubble parameter, 
and as a result, $\left(\rho_\mathrm{c} + p_\mathrm{c}\right)/\dot{H}$ has the following dependency on $R_\mathrm{h}$ irrespective of the evolution of the scale factor, namely,
\begin{align}
\frac{\rho_\mathrm{c} + p_\mathrm{c}}{\dot{H}} = \frac{\alpha(n-1)}{4\pi}\left(\frac{1}{{R_\mathrm{h}}^2}\right)\, ,
\label{II-3}
\end{align}
However, as we will see in the next subsection that this is unlike the case of $F(R)$ gravity where $\left(\rho_\mathrm{c} + p_\mathrm{c}\right)/\dot{H}$ 
indeed depends on $\dot{H}$ and thus its dependency on $R_\mathrm{h}$ relies on the specific evolution of the scale factor. 
Having obtained $\left(\rho_\mathrm{c} + p_\mathrm{c}\right)/\dot{H}$ solely in terms of $R_\mathrm{h}$, we now can perform the integration of Eq.~(\ref{MGT-8}). 
Thereby plugging the above expression of Eq.~(\ref{II-3}) into Eq.~(\ref{MGT-8}) and performing the required integration, 
we finally land with the following entropy corresponding to the $(n+1)$ dimensional GB gravity:
\begin{align}
S_\mathrm{g} =  \frac{A}{4}\left\{1 + \frac{2\alpha}{{R_\mathrm{h}}^2}\left(\frac{n-1}{n-3}\right)\right\}\, .
\label{II-4}
\end{align}
This resembles $S_\mathrm{GB}$ shown in Eq.~(\ref{correct entropy-GB}), which in turn indicates that the generalized entropy obtained in Eq.~(\ref{MGT-8}) is on the right track.

\subsection{Case-III: $(3+1)$ dimensional $F(R)$ gravity}

$F(R)$ gravity is one of the important modified theories of gravity that has rich implications in various aspects of cosmology, namely in describing different cosmic eras 
of the universe and their smooth unification (see \cite{Nojiri:2010wj, Capozziello:2009nq} for review of $F(R)$ gravity). 
Thus, finding the correct horizon entropy for the $F(R)$ theory is worthwhile to investigate. 
Here we confine ourselves in $(3+1)$ dimensional spacetime and will determine the corresponding entropy in $F(R)$ gravity by using Eq.~(\ref{MGT-8}). 
 From Eq.~(\ref{ep-FR}), we determine,
\begin{align}
\frac{\rho_\mathrm{c} + p_\mathrm{c}}{\dot{H}} = 2f'(R) + 6f''(R)\left(-4H^2 + 4\dot{H} + \frac{3H\ddot{H}}{\dot{H}} 
+ \frac{\dddot{H}}{\dot{H}}\right) + 36f'''(R)\dot{H}\left(4H + \frac{\ddot{H}}{\dot{H}}\right)^2\, ,
\label{III-1}
\end{align}
(recall that $f(R) = F(R) - R$). 
$\left(\rho_\mathrm{c} + p_\mathrm{c}\right)/\dot{H}$ depends on the derivatives of the Hubble parameter, and thus the dependency 
of $\left(\rho_\mathrm{c} + p_\mathrm{c}\right)/\dot{H}$ on $R_\mathrm{h}$ can not be obtained without specifying some specific form of the scale factor. 
Hence we consider a power law scale factor, in particular, $a(t) \propto t^{h_0}$ which, depending on the value of $h_0$, can describe different phases of the standard Big-Bang universe. 
For instance, $h_0 = \frac{2}{3}$ leads to the matter dominated universe while $h_0 = \frac{1}{2}$ gives the radiation era. 
Here we would like to mention that such power law scale factor naturally arises as one of the possible cosmological solutions in $F(R) \propto R^m$ gravity theory 
described by the action \cite{Nojiri:2010wj},
\begin{align}
\mathcal{Z} =\int d^4x \sqrt{-g}\left[f_0 R^m\right] + \mathcal{Z}_\mathrm{mat}\, ,
\label{III-1-2}
\end{align}
where $\mathcal{Z}_\mathrm{mat}$ represents some perfect fluid distribution having constant EoS.
Therefore with the above $F(R)$, we may consider the following scale factor of the universe:
\begin{align}
a(t) \propto t^{h_0} \quad \mbox{with} \quad h_0 = \frac{2m}{3(1+\omega)}\, ,
\label{III-2}
\end{align}
where $\omega$ is the constant EoS of the perfect fluid under consideration. 
The above scale factor immediately leads to the Hubble parameter, and consequently, the apparent horizon as,
\begin{align}
H(t) = \frac{h_0}{t} \quad \Longrightarrow \quad R_\mathrm{h}(t) = \frac{t}{h_0}\, ,
\label{III-3}
\end{align}
on inverting which, we get the cosmic time as a function of $R_\mathrm{h}$: $t(R_\mathrm{h}) = h_0R_\mathrm{h}$. 
As a consequence along with $a(t) \propto t^{h_0}$, the $F(R) = f_0R^m$ can be written as,
\begin{align}
F(R) = 6^mf_0\left(2 - \frac{1}{h_0}\right)^{m}\left(\frac{1}{R_\mathrm{h}}\right)^{2m}\, ,
\label{III-3-4}
\end{align}
Moreover using the above mentioned relation of $t = t(R_\mathrm{h})$ into Eq.~(\ref{III-1}), we obtain $\left(\rho_\mathrm{c} + p_\mathrm{c}\right)/\dot{H}$ solely in terms of $R_\mathrm{h}$:
\begin{align}
\frac{\rho_\mathrm{c} + p_\mathrm{c}}{\dot{H}} = -2 - \frac{\beta}{{R_\mathrm{h}}^{2m-2}} \quad \mbox{with} \quad 
\beta = \frac{mf_06^m}{3h_0}\left(2m^2 + (-3+h_0)m + 1 - 2h_0\right)\left(2 - \frac{1}{h_0}\right)^{m-1}\, ,
\label{III-4}
\end{align}
Plugging the above expression into Eq.~(\ref{MGT-8}) along with a bit of integration over $R_\mathrm{h}$ yields to the following form of entropy 
(note, $\Omega_n = 4/(3\pi)$ in $(3+1)$ dimension):
\begin{align}
S_\mathrm{g} = \frac{A}{4}\left[1 - 8\pi + \left(\frac{4\pi\beta}{m-2}\right)\frac{1}{{R_\mathrm{h}}^{2m-2}}\right]\, ,
\label{III-5}
\end{align}
which, due to Eq.~(\ref{III-3-4}), can be written as,
\begin{align}
S_\mathrm{g} = \frac{A}{4}\left[1 + 8\pi\left\{F'(R)\left(1 + \frac{(m-1)(2m-1)}{h_0(m-2)}\right) - 1\right\}\right]\, ,
\label{III-5}
\end{align}
Thus as a whole, Eq.~(\ref{III-5}) is the entropy that links the FLRW equations of action (\ref{III-1-2}) with the thermodynamic law (\ref{law-1}). 
The entropy gets a correction factor proportional to $F'(R)$ over the Bekenstein-Hawking term, and clearly, the correction factor reduces to zero for $F(R) = R$ 
(or equivalently, $f_0 = m = 1$), as expected.

\subsubsection*{Equivalence/ Non-equivalence between $F(R)$ and scalar-tensor frames from horizon entropy}

The $F(R)$ action (\ref{III-1-2}) can be mapped into the Einstein frame by applying the following conformal transformation 
on the metric $g_{\mu\nu}$ (see \cite{Nojiri:2010wj, Capozziello:2009nq} for such mapping):
\begin{equation}
g_{\mu\nu} \longrightarrow \widetilde{g}_{\mu\nu} = \e^{-\sqrt{2/3}\,\varphi}g_{\mu\nu}
\label{con transformation}\, ,
\end{equation}
where $\varphi$ is the conformal factor which is related to the spacetime curvature as $F'(R) = \e^{-\sqrt{2/3}\,\varphi}$.
If $R$ and $\widetilde{R}$ are the Ricci scalars formed by $g_{\mu\nu}$ and $\widetilde{g}_{\mu\nu}$ respectively, then they are related by the following
expression:
\begin{align}
 R = \e^{-\sqrt{2/3}\,\varphi}\left[\widetilde{R} - \widetilde{g}^{\mu\nu}\partial_{\mu}\varphi\partial_{\nu}\varphi
 + \sqrt{6}\widetilde{\Box}\varphi\right]\, ,
 \nonumber
\end{align}
with $\widetilde{\Box}$ is the d'Alembertian operator corresponding to the tilde frame. Using the above expression along with the aforementioned
relation between $\varphi$ and $F'(R)$, we get the following scalar-tensor action in Einstein frame:
\begin{align}
S = \int d^4x\sqrt{-\widetilde{g}} \left\{\frac{\widetilde{R}}{2} - \frac{1}{2}\tilde{g}^{\mu\nu}(\partial_{\mu}\varphi)(\partial_{\nu}\varphi) - V(\varphi) 
+ \e^{\frac{4}{\sqrt{6}}\varphi}\mathcal{L}_\mathrm{mat}[\rho,p]\right\}\, ,
\label{Action Einstein frame}
\end{align}
It may be observed that $\varphi$ acts as a scalar field with the potential given by \cite{Nojiri:2010wj,Capozziello:2009nq},
\begin{equation}
V(\varphi) \equiv \frac{RF'(R) - F(R)}{2F'(R)^{2}}\, ,
\label{Eq: Original potential}
\end{equation}
where, $\varphi = \varphi(R)$ obtained by the relation $\e^{-\sqrt{2/3} \varphi} = F'(R)$. 
Thus the scalar field potential explicitly depends on the form of $F(R)$. For our considered form of $F(R) = f_0R^{m}$ in Eq.~(\ref{III-1-2}), 
the corresponding scalar potential takes the following form,
\begin{align}
V(\varphi) = \left[\frac{m-1}{2(f_0 m^m)^{1/(m-1)}}\right]
\exp\left\{\left(\frac{m-2}{m-1}\right)\sqrt{\frac{2}{3}} \varphi\right\}\, ,
\label{potential 2}
\end{align}
Thus as a whole, the $F(R)$ frame can be equivalently mapped to a scalar-tensor theory by a conformal transformation of the spacetime metric, 
where the scalar potential in the scalar-tensor frame depends on the form of $F(R)$. 
Due to Eq.~(\ref{con transformation}), the proper time in the Jordan frame (symbolized by $t$) is connected to that of in the Einstein frame ($\tilde{t}$) by 
$dt = \e^{\frac{1}{2}\sqrt{\frac{2}{3}}\varphi} d\tilde{t}$, i.e., the proper times of Jordan and Einstein frame observers are not the same. 
Thus despite the mathematical connection between these two frames, the important question that may arise is whether the frames are physically equivalent. 
Here we are mainly interested in examining the physical equivalence of Jordan and the Einstein frames in entropic cosmology. 
For this purpose, we intend to determine the horizon entropy for the Einstein frame corresponding to $F(R) = f_0R^m$, and then 
we will compare it with Eq.~(\ref{III-5}). Hereafter, we use a tilde to represent a quantity in the Einstein frame.

Clearly the scalar-tensor action (\ref{Action Einstein frame}) is described by Einstein's gravity with the matter fields given by $\rho$, $p$, and $\varphi$. 
Therefore a Bekenstein-Hawking-like entropy, in particular,
\begin{align}
\widetilde{S}_\mathrm{g} = \frac{\widetilde{A}}{4} = \frac{\pi}{\widetilde{H}^2}\, ,
\label{III-6}
\end{align}
stands to be the correct horizon entropy that produces the FLRW equations for the action (\ref{Action Einstein frame}) from the horizon thermodynamics. 
Here $\widetilde{A}$ is the horizon area of the Einstein frame with $\widetilde{H}$ being the corresponding Hubble parameter. 
We need the relation between $\widetilde{H}$ and $H$ to compare the two frames from the standpoint of horizon entropy. 
The spacetime metric in the Einstein frame can be obtained from the corresponding Jordan frame by using the conformal transformation as in Eq.~(\ref{con transformation}). 
Thus the line element in the Einstein frame is given by,
\begin{align}
d \tilde{s}^2=&\, \e^{-\sqrt{2/3}\varphi}\left[-dt^2 + a^2(t)\delta_{ij}dx^{i}dx^{j}\right]\nonumber\\
=&\, -d \tilde{t}^2 + \tilde{a}^2(\tilde{t})\delta_{ij}dx^{i}dx^{j}\, ,
\label{metric anstaz Einstein frame}
\end{align}
The quantity $d\tilde{t}$ and $\tilde{a}(\tilde{t})$ represent the proper time and the proper scale factor of the universe in the Einstein frame respectively, 
and they are related to that of in the Jordan frame as,
\begin{align}
d\tilde{t} = \e^{-\sqrt{1/6}\,\varphi}dt\, ,\quad \tilde{a}(\tilde{t}) = \e^{-\sqrt{1/6}\,\varphi}~a(t)\, ,
\label{relation-time-scale factor}
\end{align}
respectively. Consequently, the Hubble parameter in the Einstein frame is defined by $\widetilde{H} = \frac{d\ln{\tilde{a}}}{d\tilde{t}}$. 
Using this definition of $\widetilde{H}$ along with Eq.~(\ref{relation-time-scale factor}), we determine how the Hubble parameter 
in the Einstein frame is related to that of in the Jordan frame, and it comes as follows:
\begin{align}
H = \e^{-\sqrt{1/6}\,\varphi}\left\{\widetilde{H} + \frac{1}{\sqrt{6}}\frac{d\varphi}{d\tilde{t}}\right\}\, ,
\label{relation Hubble parameter}
\end{align}
At this stage in the scalar-tensor case, we use our considered form of $F(R) = f_0R^m$. 
Therefore by using Eq.~(\ref{III-3-4}) along with $F'(R) = \e^{-\sqrt{2/3}\,\varphi}$, one may write $\varphi$ in terms of $H$ as,
\begin{align}
\e^{-\sqrt{2/3}\,\varphi} = mf_0\left(12 - \frac{6}{h_0}\right)^{m-1}H^{2m-2}\, ,
\label{III-7}
\end{align}
where we use $R_\mathrm{h} = 1/H$. 
Taking derivative on both sides of Eq.~(\ref{III-7}) with respect to $\frac{d}{d\tilde{t}}$ and using Eq.~(\ref{relation-time-scale factor}), to get
\begin{align}
\frac{1}{\sqrt{6}}\frac{d\varphi}{d\tilde{t}} = \frac{(m-1)}{h_0\sqrt{mf_0}\left(12 - \frac{6}{h_0}\right)^{(m-1)/2}}\left(\frac{1}{H^{m-2}}\right)\, ,
\label{III-8}
\end{align}
Plugging the above two expressions into Eq.~(\ref{relation Hubble parameter}) and with a little bit of simplification relates 
the Hubble parameter of scalar-tensor frame with the $F(R)$ Hubble parameter as,
\begin{align}
\widetilde{H} = \left(\frac{1}{\sqrt{mf_0}\left(12 - \frac{6}{h_0}\right)^{(m-1)/2}}\right)\left\{1 - \frac{m-1}{h_0}\right\}\left(\frac{1}{H^{m-2}}\right)\, ,
\label{III-9}
\end{align}
due to which, the horizon entropy in the scalar-tensor frame from Eq.~(\ref{III-6}) becomes,
\begin{align}
\widetilde{S}_\mathrm{g} = \frac{\e^{-\sqrt{2/3}\,\varphi}}{\left(1 - \frac{m-1}{h_0}\right)^2}\left(\frac{A}{4}\right) \equiv
\frac{F'(R)}{\left(1 - \frac{m-1}{h_0}\right)^2}\left(\frac{A}{4}\right)\, ,
\label{III-10}
\end{align}
Here we use Eq.~(\ref{III-7}) and $A$ (without tilde) is the horizon area in $F(R)$ frame, moreover the second equality comes from $F'(R) = \e^{-\sqrt{2/3}\,\varphi}$. 
It is clear from Eq.~(\ref{III-5}) and Eq.~(\ref{III-10}) that the horizon entropy in the Jordan frame does not match that of in Einstein frame. 
This seems to spoil the equivalence between Jordan and Einstein frames from the perspective of horizon entropy that produces the FLRW equations 
of the respective frames from the apparent horizon thermodynamics.

\section{Problem with the existing thermodynamic law (\ref{law-1}) and a new thermodynamic law for cosmology:}\label{sec-problem}

It would be nice if this becomes the end of the story. However, unfortunately, it turns out that the thermodynamic law (\ref{law-1}) faces serious difficulties,
namely --- 
(a) the thermodynamic law $TdS = -dE + WdV$ leads to an inconsistency for $\omega \neq -1$, as demonstrated in \cite{Sanchez:2022xfh}, and (b) the entropy does not exist for $\left(1 + \frac{\dot{H}}{2H^2}\right) < 0$ (as already discussed in the last paragraph of Sec.~[\ref{law-1-Einstein}]). Following \cite{Sanchez:2022xfh}, here we briefly discuss the first problem in $(n+1)$ dimensional spacetime, and then, we propose a new thermodynamic law for cosmological apparent horizon in order to resolve the issue. Note, the authors of \cite{Sanchez:2022xfh} considered a $(3+1)$ dimensional spacetime, however we will generalize it in higher dimension and will show that the problem persists even in $(n+1)$ dimension. The basis of Sec.~[\ref{law-1-Einstein}] is to start from the thermodynamic law (\ref{law-1}), and based on this law, we have derived the horizon entropy corresponding to the Friedmann equations. In this regard, one may check the consistency by considering a bottom-up approach, in particular, one can start from the horizon entropy itself and then try to obtain the underlying thermodynamic law to examine whether the obtained thermodynamic law matches with the original one or not. Clearly, for a self consistent scenario, these two approaches will lead to the same result. However it turns out that for the thermodynamic law (\ref{law-1}), these two direct and reverse approaches do not give the same picture, unless $\omega = -1$; and the demonstration goes as follows: due to the Friedmann Eq.(\ref{FRW-Einstein}), the Bekenstein-Hawking entropy can be written by,
\begin{align}
 S = \frac{n}{4}\Omega_\mathrm{n}\left[\frac{n(n-1)}{16\pi \rho}\right]^{(n-1)/2}\, ,
 \label{prob-R-1}
\end{align}
which, owing to $\rho = E/V$, becomes,
\begin{align}
 S = \frac{n}{4}\Omega_\mathrm{n}\left[\sqrt{\frac{n(n-1)}{16\pi}\left(\frac{V}{E}\right)}\right]^{n-1}\, .
 \label{prob-R-2}
\end{align}
Eq.~(\ref{prob-R-2}) gives a relation between three extensive variables, in particular, $S = S(E,V)$. This immediately provides the total differential of $S$ as,
\begin{align}
 dS = \left(\frac{\partial S}{\partial E}\right)dE + \left(\frac{\partial S}{\partial V}\right)dV\, ,
 \label{prob-R-3}
\end{align}
which, due to Eq.~(\ref{prob-R-2}), takes the following form,
\begin{align}
 \left(\frac{H}{2\pi}\right)dS = -dE + \rho dV\, .
 \label{prob-R-4}
\end{align}
Comparing the above expression with Eq.~(\ref{law-1}), we may argue that both the expressions do not coincide unless $\rho + p = 0$ which represents an EoS as $\omega = -1$. Therefore the thermodynamic law (\ref{law-1}) leads to an inconsistent picture for $\omega \neq -1$, in particular --- starting from Eq.~(\ref{law-1}), we get the Bekenstein-Hawking entropy as the correct horizon entropy corresponding to the FLRW Eq.~(\ref{FRW-Einstein}) (see Sec.~[\ref{law-1-Einstein}]), however starting from the Bekenstein-Hawking entropy itself, one lands to Eq.~(\ref{prob-R-4}) which is different than the original thermodynamic law $TdS = -dE + WdV$ for $\omega \neq -1$. This indicates that the thermodynamic law of Eq.~(\ref{law-1}), i.e., $TdS = -dE + WdV$ is consistent only for the special case $\omega = -1$. Therefore in order to have a thermodynamic law that holds for all values of EoS of the matter field, we propose a modified thermodynamic law in the context of cosmology, as follows:
\begin{align}
TdS = -dE + \rho dV\, ,
\label{prob-4}
\end{align}
where $T$ is shown in Eq.~(\ref{AH2}). At some stage in the next section, we will show that such modified thermodynamic law resolves the aforementioned problem for all possible values of the EoS of the matter field and thus is considered to be more general compared to the previous one (\ref{law-1}) which, however, is a limiting case of the modified thermodynamics for $p = -\rho$.\\

This modified thermodynamic law surely affects the horizon entropy (compared to the previously found in Eq.~(\ref{MGT-8})) that links thermodynamics
with FLRW equations in a general gravity theory.
Therefore the immediate task is to find the cosmic entropy associated with the newly proposed law (\ref{prob-4}) in a general gravitational theory.
This is the subject of the next subsections where we will also see that, besides the above-mentioned resolution, the entropy based on (\ref{prob-4})
will remain valid irrespective of whether $\left(1 + \frac{\dot{H}}{2H^2}\right)$ is positive or negative, unlike the case of (\ref{law-1})
where the entropy does not exist when $\left(1 + \frac{\dot{H}}{2H^2}\right)$ becomes negative.
Thus the thermodynamic relation (\ref{prob-4}) proves to simultaneously resolve both the difficulties of (\ref{law-1}).
%Here we would like to mention that whatever the form of the modified entropy will be, it will reduce to the older one (\ref{MGT-8})
%for $p_\mathrm{eff} = -\rho_\mathrm{eff}$ as the new thermodynamic law (\ref{prob-4}) itself converges to the older one (\ref{law-1})
%for $p_\mathrm{eff} = -\rho_\mathrm{eff}$.

Before going to any modified theories of gravity, we will first discuss the case of Einstein's gravity.

\subsection{Consistent entropy for Einstein's gravity with the modified thermodynamic law}

Let us first recall that the $(n+1)$ dimensional FLRW equations in Einstein's gravity are given in Eq.~(\ref{FRW-Einstein}). 
With $E = \rho V$, Eq.~(\ref{prob-4}) can be written as,
\begin{align}
\frac{H}{2\pi}\left|1 + \frac{\dot{H}}{2H^2}\right|\frac{dS_\mathrm{E}^{(m)}}{dt} = -\dot{\rho}V\, ,
\label{E-L-1}
\end{align}
Due to the conservation of matter fields in Eq.~(\ref{conservation law}) along with the FLRW equations, Eq.~(\ref{E-L-1}) turns out to be (see the Appendix in Sec.~[\ref{sec-appendix}]),
\begin{align}
\frac{dS_\mathrm{E}^{(m)}}{dt} = \frac{n(n-1)}{4}\Omega_n\left(\frac{{R_\mathrm{h}}^{n-2}\dot{R}_\mathrm{h}}{\left|1 + \frac{\dot{H}}{2H^2}\right|}\right)\, ,
\label{E-L-2}
\end{align}
(where the suffix `E' is for Einstein's gravity and the superfix `$m$' is because of we are working with the modified thermodynamic law) on integrating which,
we obtain the final form of the required entropy as follows,
\begin{align}
S_\mathrm{E}^{(m)} = \frac{n(n-1)}{4}\Omega_n\int \frac{{R_\mathrm{h}}^{n-2}}{\left|1 + \frac{\dot{H}}{2H^2}\right|}dR_\mathrm{h}\, ,
\label{E-L-3}
\end{align}
At this stage, we may recall our argument made in the previous section, in particular, the entropy based on the modified thermodynamic law is valid irrespective 
of whether $\left(1 + \frac{\dot{H}}{2H^2}\right)$ is positive or negative. 
This is evident from the above equation. 
Therefore the modified thermodynamic law can bridge the underlying thermodynamics with the FLRW equations of Einstein's gravity even for $\left(1 + \frac{\dot{H}}{2H^2}\right)<0$, 
or equivalently, for $\omega > \frac{1}{3}$ (which may occur during the scalar field-dominated reheating stage with suitable scalar potential). 
This is one of the main advantages of the modified thermodynamic law over the older one (\ref{law-1}). 
Moreover, Eq.~(\ref{E-L-3}) also indicates that the integral (over $R_\mathrm{h}$) depends on the derivative of the Hubble parameter 
and thus can be realized for a certain cosmological evolution of the universe, i.e., for a certain scale factor $a = a(t)$. 
However, for a constant EoS of the matter fields (for instance, if the matter field is dominated by some perfect fluid), 
the above integral can be performed without specifying any particular form of the scale factor. 
This is because the factor $\left(1 + \frac{\dot{H}}{2H^2}\right) = 1-\frac{n}{4}\left(1+\omega\right)$ from the FLRW Eq.~(\ref{FRW-Einstein}), 
which becomes constant for a constant $\omega = p/\rho$. Thus for a constant $\omega$, the term containing $\dot{H}$ can be taken outside of the integral, and consequently, 
the entropy from Eq.~(\ref{E-L-3}) turns out to be,
\begin{align}
S_\mathrm{E}^{(m)}(\mathrm{constant~\omega}) = \frac{A}{\left|4 - n - n\omega\right|}\, ,
\label{E-L-4}
\end{align}
Having obtained the entropy, we now will examine whether the modified thermodynamic law (\ref{prob-4}) can resolve the first problem stated in the beginning of Sec.~[\ref{sec-problem}]. Note that to get Eq.~(\ref{E-L-4}), we have considered the modified thermodynamic law (\ref{prob-4}), and based on this law, we have derived the corresponding horizon entropy such that the cosmological field equations get satisfied. However, now we will test the bottom-up approach, in particular, we will start from the entropy of Eq.~(\ref{E-L-4}) and then try to achieve the underlying thermodynamic law. For this purpose, let us consider the Friedmann Eq.(\ref{FRW-Einstein}), due to which, the entropy in Eq.~(\ref{E-L-4}) can be written as,
\begin{align}
 S_\mathrm{E}^{(m)} = \frac{n}{\left|4-n-n\omega\right|}\Omega_\mathrm{n}\left[\frac{n(n-1)}{16\pi \rho}\right]^{(n-1)/2}\, .
 \label{prob-M-1}
\end{align}
Owing to $\rho = E/V$, the above expression becomes,
\begin{align}
 S_\mathrm{E}^{(m)} = \frac{n}{\left|4-n-n\omega\right|}\Omega_\mathrm{n}\left[\sqrt{\frac{n(n-1)}{16\pi}\left(\frac{V}{E}\right)}\right]^{n-1}\, .
 \label{prob-M-2}
\end{align}
The above relation of $S_\mathrm{E}^{(m)} = S_\mathrm{E}^{(m)}(E,V)$ immediately provides the total differential of $S_\mathrm{E}^{(m)}$ as follows,
\begin{align}
 \left(\frac{H}{2\pi}\right)\left|1 - \frac{n}{4}\left(1+\omega\right)\right|dS_\mathrm{E}^{(m)} = -dE + \rho dV\, ,
 \label{prob-M-3}
\end{align}
which, because of $\left(1 + \frac{\dot{H}}{2H^2}\right) = 1-\frac{n}{4}\left(1+\omega\right)$, can be equivalently given by,
\begin{align}
 \left(\frac{H}{2\pi}\right)\left|1 + \frac{\dot{H}}{2H^2}\right|dS_\mathrm{E}^{(m)} = -dE + \rho dV\, .
 \label{prob-M-4}
\end{align}
Clearly the above expression matches with the modified thermodynamic law (\ref{prob-4}) by identifying the $T = \left(\frac{H}{2\pi}\right)\left|1 + \frac{\dot{H}}{2H^2}\right|$ from Eq.(\ref{AH2}). Therefore for the two approaches given by --- whether we start from thermodynamic law and then derive the corresponding horizon entropy or we start from the horizon entropy and then achieve the underlying thermodynamic law; the modified thermodynamic law in Eq.~(\ref{prob-4}) leads to consistent picture for all possible values of $\omega$, unlike to the previous thermodynamic law in Eq.~(\ref{law-1}) which results to an inconsistency in these two approaches for $\omega \neq -1$. Thus the modified thermodynamic law proves to be consistent for all values of EoS of the matter fields.\\

Coming back to the entropy of Eq.~(\ref{E-L-4}), it explicitly depends on the value of $\omega$.
Thereby in this modified thermodynamic law, the form of the entropy changes with the evolution era of the universe, for instance,
\begin{align}
\begin{array}{ll}
S_\mathrm{E}^{(m)}(\mathrm{constant~\omega})= \frac{A}{4}\, , &
\mbox{during inflation when}\ \omega =\, -1\, ,\\
S_\mathrm{E}^{(m)}(\mathrm{constant~\omega})= \frac{A}{\left|4-n\right|}\, ,&
\mbox{during matter-dominated era when}\ \omega = 0\, ,\\
S_\mathrm{E}^{(m)}(\mathrm{constant~\omega})= \frac{A}{4\left|1-n/3\right|}\, ,& 
\mbox{during radiation era when}\ \omega = 1/3\, ,
\end{array}
\nonumber
\end{align}
Several points to be noted ---
(a) for $\omega=-1$, the entropy coming from the modified thermodynamic law (\ref{prob-4}) matches with that of coming from the previous thermodynamic law (\ref{law-1}). 
This is actually expected because the relation (\ref{prob-4}) itself converges to Eq.~(\ref{law-1}) for $\omega = -1$. 
(b) For $\omega = 1/3$ and $n = 3$ (i.e., in $(3+1)$ dimension), the entropy becomes ill-defined. 
This is because the horizon temperature vanishes in these limits (see Eq.~(\ref{AH2})), in which case, a proper interpretation of horizon entropy requires certain modifications. (c) The entropy of the apparent horizon in cosmology resembles with the Schwarzschild black hole entropy only for $\omega = -1$. This is because the fact that the cosmological apparent horizon becomes static for $\omega = -1$, which is actually similar with the black hole system. However for $\omega \neq -1$, the cosmological apparent horizon gets dynamic in nature, unlike a black hole where the event horizon is static. As a consequence, the horizon entropy in cosmology for $\omega \neq -1$ differs from that of a black hole.

Besides the constant $\omega$ case, if the EoS of the matter field varies with the cosmic evolution, then the entropy from Eq.~(\ref{E-L-3}) is given by,
\begin{align}
S_\mathrm{E}^{(m)}(\mathrm{variable~\omega}) = \frac{n(n-1)}{4}\Omega_n\int \frac{{R_\mathrm{h}}^{n-2}}{\left|4 - n - n\omega(R_\mathrm{h})\right|}dR_\mathrm{h}\, ,
\label{E-L-5}
\end{align}
which can not be realized without specifying a certain cosmological scale factor.

Thus as a whole, Eq.~(\ref{E-L-3}) provides the entropy that connects the FLRW Eq.~(\ref{FRW-Einstein}) with the modified thermodynamic law (\ref{prob-4}). 
In this scenario, the entropy seems to depend on the EoS of the matter field, where Eq.~(\ref{E-L-4}) and Eq.~(\ref{E-L-5}) show 
the cases for constant $\omega$ and variable $\omega$, respectively.

\subsection{Consistent entropy for general gravity theory with the modified thermodynamic law}\label{sec-law-2-MGT}

The action for a general gravity theory in $(n+1)$ dimension is shown in Eq.~(\ref{MGT-action}) and the corresponding FLRW equations are given by Eq.~(\ref{FRW-MGT}). 
In this case, the possible effect(s) of higher order curvature term(s) are encoded within $\rho_\mathrm{c}$ and $p_\mathrm{c}$ in the FLRW equations. By following the analogy stated before Eq.~(\ref{correct entropy-MGT-1}), the horizon entropy corresponding to a general modified gravity theory is considered to have the following form, namely,
\begin{align}
S_\mathrm{g}^{(m)} = \frac{n(n-1)}{4}\Omega_n\int \frac{{R_\mathrm{h}}^{n-2}}{\left|1 + \frac{\dot{H}}{2H^2}\right|}dR_\mathrm{h} + S_\mathrm{c}^{(m)}(A)\, ,
\label{MGT-L-1}
\end{align}
where $S_\mathrm{c}^{(m)}(A)$ is a non-trivial function of the area of the apparent horizon, which we want to find such that the modified thermodynamic relation:
\begin{align}
T\frac{dS_\mathrm{g}^{(m)}}{dt} = -\frac{dE}{dt} + \rho\frac{dV}{dt}\, ,
\label{MGT-L-2}
\end{align}
 holds. From Eq.~(\ref{MGT-L-1}), the above relation can be written as,
\begin{align}
T\frac{dS_\mathrm{c}^{(m)}}{dt} = \frac{d}{dt}\left(\rho_\mathrm{c}V\right) - \rho_\mathrm{c}\dot{V} + \left[-\frac{d}{dt}\left(\rho V + \rho_\mathrm{c}V\right)
+ \left(\rho + \rho_\mathrm{c}\right)\dot{V} - \left(\frac{H}{2\pi}\right)\frac{n(n-1)}{4}\Omega_n {R_\mathrm{h}}^{n-2}\dot{R}_\mathrm{h}\right]\, ,
\label{MGT-L-3}
\end{align}
where we use Eq.~(\ref{AH2}). Owing to the FLRW Eq.~(\ref{FRW-MGT}) along with $\nabla_{\mu}Q^{\mu\nu} = 0$, one may show that the terms within the square bracket of Eq.~(\ref{MGT-L-3}) cancel each other (see the Appendix of Sec.~[\ref{sec-appendix}]).
As a result, Eq.~(\ref{MGT-L-3}) becomes,
\begin{align}
T\frac{dS_\mathrm{c}^{(m)}}{dt} = \frac{d}{dt}\left(\rho_\mathrm{c}V\right) - \rho_\mathrm{c}\dot{V} = \dot{\rho}_\mathrm{c}V\, ,
\label{MGT-L-4}
\end{align}
which, due to the extended Bianchi identity: $\nabla_{\mu}Q^{\mu\nu} = 0$, turns out to be,
\begin{align}
T\frac{dS_\mathrm{c}^{(m)}}{dt} = -nH\left(\rho_\mathrm{c} + p_\mathrm{c}\right)V\, ,
\label{MGT-L-5}
\end{align}
The above equation is a single order differential equation for $S_\mathrm{c}^{(m)}$ which, by considering Eq.~(\ref{AH2}), can be integrated to get,
\begin{align}
S_\mathrm{c}^{(m)} = 2\pi n\Omega_n\int {R_\mathrm{h}}^{n-2}\left(\frac{\rho_\mathrm{c} + p_\mathrm{c}}{\dot{H}\left|1 + \frac{\dot{H}}{2H^2}\right|}\right)dR_\mathrm{h}\, ,
\label{MGT-L-6}
\end{align}
Clearly for $\rho_\mathrm{c} = p_\mathrm{c} = 0$, i.e., for Einstein's gravity, the correction term $S_\mathrm{c}^{(m)}$ in the above equation goes to zero, as expected. 
Thus as a whole, the entropy from Eq.~(\ref{MGT-L-1}) is obtained as,
\begin{align}
S_\mathrm{g}^{(m)} = \frac{n(n-1)\Omega_n}{4}\int \frac{{R_\mathrm{h}}^{n-2}}{\left|1 + \frac{\dot{H}}{2H^2}\right|}dR_\mathrm{h} 
+ 2\pi n\Omega_n\int {R_\mathrm{h}}^{n-2}\left(\frac{\rho_\mathrm{c} + p_\mathrm{c}}{\dot{H}\left|1 + \frac{\dot{H}}{2H^2}\right|}\right)dR_\mathrm{h}\, ,
\label{MGT-L-7}
\end{align}
where we use Eq.~(\ref{E-L-3}) to get $S_\mathrm{E}^{(m)}$. Eq.~(\ref{MGT-L-7}) gives the required entropy that bridges the FLRW Eq.~(\ref{FRW-MGT}) 
of any gravity theory with the modified thermodynamic law (\ref{prob-4}), where the information of a specific gravity theory under consideration is encapsulated 
within $S_\mathrm{g}^{(m)}$ through the factor $\left(\rho_\mathrm{c} + p_\mathrm{c}\right)/\dot{H}$. 
The following points need to be mentioned in this regard --- 
(a) similar to the case of Einstein's gravity, the entropy $S_\mathrm{g}^{(m)}$ proves to exist irrespective of whether $\left(1 + \frac{\dot{H}}{2H^2}\right)$ is positive or negative; 
(b) $S_\mathrm{g}^{(m)}$ becomes ill-defined for the cosmic evolution: $\dot{H} = -2H^2$ (or equivalently $H(t) = \frac{1}{2t}$). 
Actually, in this case, the temperature of the apparent horizon also vanishes, and consequently, the horizon entropy shows a diverging behavior; and 
(c) for $\dot{H} = 0$ from the FLRW Eq.~(\ref{FRW-MGT}), and thus the entropy from Eq.~(\ref{MGT-L-7}) reduces to,
\begin{align}
S_\mathrm{g}^{(m)} = \frac{A}{4} + 2\pi n\Omega_n\int {R_\mathrm{h}}^{n-2}\left(\frac{\rho_\mathrm{c} + p_\mathrm{c}}{\dot{H}}\right)dR_\mathrm{h}\, ,
\label{MGT-L-8}
\end{align}
The comparison of Eq.~(\ref{MGT-8}) and Eq.~(\ref{MGT-L-8}) immediately leads to the argument that for $\dot{H} = 0$,
the entropies for general gravity theory coming from the two thermodynamic laws (i.e., from Eq.~(\ref{law-1}) and from Eq.~(\ref{prob-4})) match with each other.

As earlier, below we will consider some specific gravity theories and will determine the corresponding horizon entropy from Eq.~(\ref{MGT-L-7}) 
in this modified thermodynamic scenario, and will compare them with the older ones.

\subsubsection*{In absence of matter fields}

For $\rho = p = 0$, the FLRW Eq.~(\ref{FRW-MGT}) turns out to be,
\begin{align}
H^2 = \frac{16\pi}{n(n-1)}~\rho_\mathrm{c} \quad \mbox{and} \quad \dot{H} = -\frac{8\pi}{(n-1)}\left(\rho_\mathrm{c} + p_\mathrm{c}\right)\, ,
\label{I-L-1}
\end{align}
Therefore $\left(\rho_\mathrm{c} + p_\mathrm{c}\right)/\dot{H} = -(n-1)/(8\pi)$. By using which, $S_\mathrm{c}^{(m)}$ from Eq.~(\ref{MGT-L-6}) boils down to,
\begin{align}
S_\mathrm{c}^{(m)} = -\frac{n(n-1)\Omega_n}{4}\int \frac{{R_\mathrm{h}}^{n-2}}{\left|1 + \frac{\dot{H}}{2H^2}\right|}dR_\mathrm{h}\, ,
\label{I-L-2}
\end{align}
by plugging which into Eq.~(\ref{MGT-L-7}), immediately results to,
\begin{align}
 S_\mathrm{g}^{(m)} = 0~~(\textrm{or a constant})\, .
 \label{I-L-3}
\end{align}
Therefore in the modified thermodynamic law, the entropy for any gravity theory $without$ matter fields turns out to be constant.
This incident is similar compared to that of in the previous thermodynamic law, see Sec.~[\ref{sec-case-I}].

\subsubsection*{$(n+1)$ dimensional Gauss-Bonnet gravity}

The $(n+1)$ dimensional Gauss-Bonnet (GB) gravitational action is shown in Eq.~(\ref{GB-action}), for which, 
we have determined $\left(\rho_\mathrm{c} + p_\mathrm{c}\right)/\dot{H} = \frac{\alpha(n-1)}{4\pi}\left(\frac{1}{{R_\mathrm{h}}^2}\right)$, 
where $\alpha$ is the GB parameter, see Eq.~(\ref{II-2}). Therefore the entropy from Eq.~(\ref{MGT-L-7}) becomes,
\begin{align}
S_\mathrm{g}^{(m)} = \frac{n(n-1)\Omega_n}{4}\int \frac{{R_\mathrm{h}}^{n-2}}{\left|1 + \frac{\dot{H}}{2H^2}\right|}dR_\mathrm{h} 
+ \frac{\alpha n(n-1)}{2}~\Omega_n\int \frac{R_\mathrm{h}^{n-4}}{\left|1 + \frac{\dot{H}}{2H^2}\right|}dR_\mathrm{h}\, ,
\label{II-L-1}
\end{align}
Being the integral depends on the derivative of the Hubble parameter, we will consider two specific cosmological evolutions to perform the above integration. 
In particular, we consider,
\begin{align}
H=&\, \mathrm{constant}~~~~(\mbox{during inflation})\, ,\nonumber\\
H=&\, \frac{h_0}{t}~~~~~~~~~~~~(\mbox{during SBBC})\, ,
\label{II-L-2}
\end{align}
with $h_0$ be a constant (and SBBC stands as the Standard Big-Bang Cosmology).
Such evolutions in GB gravity can be realized with suitable forms of $\rho$ and $p$ in the FLRW equations. 
For instance, $H = \mathrm{constant}$ can be obtained from $p = -\rho$ even in the $(n+1)$ dimensional GB gravity, see Eq.~(\ref{FRW-GB}). 
Moreover the power law evolution, i.e., $H = h_0/t$, in the GB gravity is possible from following $\rho$ and $p$:
\begin{align}
\rho = \frac{n(n-1)}{16\pi}\left(\frac{h_0^2}{t^2} + \frac{\alpha h_0^4}{t^4}\right) \quad \mbox{and} \quad p=\frac{(n-1)}{8\pi}\left(\frac{h_0}{t^2} + \frac{2\alpha h_0^3}{t^4}\right)\, ,
\label{II-L-3}
\end{align}
respectively. 
Therefore a suitable time-dependent EoS of the matter field given by Eq.~(\ref{II-L-3}) results in a power law evolution in the GB gravity. 
Eq.~(\ref{II-L-3}) clearly indicates that the non-zero value of $\alpha$ which actually controls the deviation of the GB gravity from 
the usual Einstein's gravity makes the EoS time-dependent, otherwise $\frac{p}{\rho}$ becomes constant for $\alpha = 0$.

Having considered specific forms of the Hubble parameter, we now perform the integration of Eq.~(\ref{II-L-1}). By doing so, we obtain the required entropy from Eq.~(\ref{II-L-1}) as follows, 
\begin{align}
S_\mathrm{g}^{(m)} = \frac{A}{4}\left\{1 + \frac{2\alpha}{{R_\mathrm{h}}^2}\left(\frac{n-1}{n-3}\right)\right\}\, ,
\label{II-L-4}
\end{align}
for $H = \mathrm{constant}$, and
\begin{align}
S_\mathrm{g}^{(m)} = \left(\frac{1}{\left|1 - 1/(2h_0)\right|}\right)\frac{A}{4}\left\{1 + \frac{2\alpha}{{R_\mathrm{h}}^2}\left(\frac{n-1}{n-3}\right)\right\}\, ,
\label{II-L-5}
\end{align}
for $H = h_0/t$, with $A = n\Omega_n{R_\mathrm{h}}^{n-1}$ is the area of the apparent horizon. 
Therefore Eq.~(\ref{II-L-4}) and Eq.~(\ref{II-L-5}) provide the entropy from the modified thermodynamic law for certain cosmic evolutions in $(n+1)$ dimensional GB gravity. 
Here we would like to mention that regarding the power law evolution, particularly for $h_0 = 1/2$, the horizon temperature vanishes, and consequently, 
the entropy $S_\mathrm{g}^{(m)}$ from Eq.~(\ref{II-L-5}) becomes ill-defined. 
Moreover for $H = \mathrm{constant}$, the $S_\mathrm{g}^{(m)}$ of Eq.~(\ref{II-L-4}) looks similar to that of
in Eq.~(\ref{II-4}) coming from the previous thermodynamic law, which indeed supports the argument around Eq.~(\ref{MGT-L-8}).

\subsubsection*{$(3+1)$ dimensional $F(R)$ gravity}

Let us start by recalling Eq.~(\ref{III-1}), namely,
\begin{align}
\frac{\rho_\mathrm{c} + p_\mathrm{c}}{\dot{H}} = 2f'(R) + 6f''(R)\left(-4H^2 + 4\dot{H} + \frac{3H\ddot{H}}{\dot{H}} + \frac{\dddot{H}}{\dot{H}}\right) 
+ 36f'''(R)\dot{H}\left(4H + \frac{\ddot{H}}{\dot{H}}\right)^2\, ,
\label{III-L-1}
\end{align}
in $(3+1)$ dimensional $F(R)$ gravity theory, where $f(R)$ being the correction of $F(R)$ action over the Einstein-Hilbert term. 
As earlier, here we again consider the power law cosmic evolution to determine $\left(\rho_\mathrm{c} + p_\mathrm{c}\right)/\dot{H}$ solely in terms of $R_\mathrm{h}$. 
In particular, we consider $H(t) = h_0/t$. 
Such kind of the Hubble parameter in $F(R)$ gravity can be achieved from the $F(R)$ action like Eq.~(\ref{III-1-2}) where $S_\mathrm{mat}$ 
denotes some perfect fluid distribution with a constant EoS. 
For such $F(R) \propto R^{m}$ along with $H(t) = h_0/t$, the above expression of $\left(\rho_\mathrm{c} + p_\mathrm{c}\right)/\dot{H}$ turns down to 
Eq.~(\ref{III-4}) as determined earlier. Using such expressions of $\left(\rho_\mathrm{c} + p_\mathrm{c}\right)/\dot{H}$ and $H(t)=h_0/t$ into 
Eq.~(\ref{MGT-L-7}), and performing the required integration over $R_\mathrm{h}$, we get the desired entropy corresponding to the $F(R)$ gravity under consideration 
as follows (note, $\Omega_n = 4/(3\pi)$ in $(3+1)$ dimension), 
\begin{align}
S_\mathrm{g}^{(m)} = \left(\frac{1}{\left|1 - 1/(2h_0)\right|}\right)\frac{A}{4}\left[1 - 8\pi + \left(\frac{4\pi\beta}{m-2}\right)\frac{1}{{R_\mathrm{h}}^{2m-2}}\right]\, ,
\label{III-L-2}
\end{align}
which, due to Eq.~(\ref{III-3-4}), can be written as,
\begin{align}
S_\mathrm{g}^{(m)} = \left(\frac{1}{\left|1 - 1/(2h_0)\right|}\right)\frac{A}{4}\left[1 + 8\pi\left\{F'(R)\left(1 + \frac{(m-1)(2m-1)}{h_0(m-2)}\right) - 1\right\}\right]\, ,
\label{III-L-3}
\end{align}
The above entropy can link the FLRW equations of the action (\ref{III-1-2}) with the modified thermodynamic law (\ref{prob-4}). 
The comparison of Eq.~(\ref{III-5}) and Eq.~(\ref{III-L-3}) leads to the argument that the entropy (for the $F(R)$ gravity) based 
on the modified thermodynamics acquires an extra constant factor given by $\left|1 - 1/(2h_0)\right|^{-1}$ with respect to 
that of based on the previous thermodynamic relation in Eq.~(\ref{law-1}). Moreover, the extra factor changes depending on the value of $h_0$, 
for instance, during the matter era of the universe when $h_0 = \frac{2}{3}$, such extra factor in the modified thermodynamic scenario provides a total contribution 
of $\left|1 - 1/(2h_0)\right|^{-1} = 3$ (see the note in \footnote{Here it may be mentioned that in the context of $F(R)$ gravity, since the higher curvature term(s) 
also contribute an energy density along with the matter fields, which is why, the ``matter era'' in the previous statement means that the $effective$ energy density 
$\rho_\mathrm{eff} = \rho + \rho_\mathrm{c}$ behaves like pressureless dust with $\omega_\mathrm{eff} = 0$.}). On the other hand, 
the presence of that factor makes the entropy divergent for $h_0 = \frac{1}{2}$ during the radiation era of the universe --- 
this is the case when the temperature of the horizon vanishes, and consequently, the entropy becomes ill-defined. Here we may recall that similar to 
the $F(R)$ gravity, the entropy in Einstein's gravity as well as in $(n+1)$ dimensional GB theory also shows a diverging behavior in the situation when the horizon temperature vanishes. 
Due to Eq.~(\ref{MGT-L-7}) and the discussion around it, we may argue that such diverging character is a generic property of horizon entropy 
in the realm of the modified thermodynamic law (\ref{prob-4}), irrespective of the gravity theory under consideration.

\section{Conclusion}

In the realm of entropic cosmology, the crucial step is to determine the horizon entropy that links the cosmological field equations for a certain gravity theory 
from the underlying thermodynamics of the apparent horizon. 
In this regard, one may borrow the entropy from the corresponding black hole thermodynamics by replacing the black hole horizon with the apparent horizon. 
For instance, the Bekenstein-Hawking-like entropy can provide the FLRW equations of Einstein's gravity from the horizon thermodynamics, 
and thus the Bekenstein-Hawking entropy is considered to be the correct horizon entropy for Einstein's gravity even in the context of cosmology. 
Moreover, in the case of $(n+1)$ dimensional Gauss-Bonnet theory, a similar form of black hole entropy proves to be the correct entropy 
for the apparent horizon in the cosmology sector. 
However, the scenario becomes different for other modified theories of gravity, like the $F(R)$ theory. In particular, the black hole-like entropy of $F(R)$ gravity 
is not able to produce the FLRW equations of $F(R)$ theory from the cosmic thermodynamic law, and thus the correct horizon entropy for $F(R)$ cosmology remains unknown. 
This naturally raises the important question --- what is the form of horizon entropy that connects the FLRW equations for a ``$general$'' gravity theory 
with the underlying thermodynamics of the apparent horizon?

In the present paper, we take this issue and determine a general form of horizon entropy that establishes the inter-relation between horizon thermodynamics 
and cosmology for any modified theories of gravity, where the thermodynamic law is given by $TdS = -dE + WdV$ (with $W = \left(\rho - p\right)/2$ represents 
the work density of the matter field and the other symbols have their usual meaning). 
By using such generalized entropy, we find the respective horizon entropies for several modified theories of gravity. As a result, we notice that for Einstein's gravity and 
the $(n+1)$ dimensional Gauss-Bonnet theory, the generalized entropy reduces to a similar form of their respective black hole entropy, however, in the case of $F(R)$ gravity, 
the generalized entropy reduces to an entirely different form than the corresponding $F(R)$ black hole entropy. It would be nice if this becomes the end of the story where, 
by using the generalized entropy, one may determine the correct entropy of the apparent horizon for any modified theories of gravity. 
However, it turns out that besides the aforementioned question, the thermodynamic law $TdS = -dE + WdV$ itself has some serious difficulties, like ---
(a) it leads to some inconsistency  for $\omega \neq -1$ (see Sec.[\ref{sec-problem}] for the details), and moreover,
(b) for $\omega > 1/3$, there exists no such horizon entropy (based on $TdS = -dE + WdV$) that links the cosmology with the apparent horizon thermodynamics (here it may be mentioned that $\omega > 1/3$ may arise during the reheating stage
in standard scalar field cosmology for suitable scalar potential).
This indicates that the thermodynamic law $TdS = -dE + WdV$ is valid only for the special case when the EoS of the matter field is $=-1$, i.e., $p = -\rho$. 
Motivated by this problematic issue, we propose a modified thermodynamic law of apparent horizon, given by $TdS = -dE + \rho dV$, which seems to be free from such difficulties. 
Therefore the modified thermodynamic law is valid for all values of EoS of the matter field and thus is considered to be more general compared to the previous one which, 
however, is a limiting case of the modified thermodynamics for $\omega = -1$. 
This modified thermodynamic law surely affects the horizon entropy (compared to that based on the previous thermodynamic law) 
that links the horizon thermodynamics with FLRW equations in a general gravity theory. 
Thus, based on such modified thermodynamics, i.e., on $TdS = -dE + \rho dV$, we further determine a generalized entropy that gives the cosmology 
of any gravity theory for all values of EoS of the matter field. 
In this case, the entropy for certain gravity theory does not look similar to that of black hole thermodynamics, even for Einstein and for $(n+1)$ dimensional GB gravity theory. 
For instance, in Einstein's gravity, the entropy based on the modified thermodynamic law gets different than the Bekenstein-Hawking entropy by an $\omega$ dependent pre-factor. 
Moreover, we would like to mention that the entropies (for certain gravity theories) coming from the two different thermodynamic laws match with each other for $\omega = -1$; 
this is because the thermodynamic laws themselves converge for $p = -\rho$.

As a whole, the present work provides the correct horizon entropy that connects the cosmological field equations of a general gravity theory (and for all EoS 
of the matter field) with the horizon thermodynamics. We hope this will be a big leap regarding the interconnection between cosmology and the underlying thermodynamics of the apparent horizon.

\section{Appendix}\label{sec-appendix}

In order to establish Eq.~(\ref{E-L-2}), we first use the conservation of matter field, namely $\dot{\rho}+ nH\left(\rho + p\right) = 0$, into Eq.~(\ref{E-L-1}).
This results to,
\begin{align}
\frac{H}{2\pi}\left|1 + \frac{\dot{H}}{2H^2}\right|\frac{dS_\mathrm{E}^{(m)}}{dt} =&\, nH\left(\rho + p\right)\Omega_n{R_\mathrm{h}}^n\, ,\nonumber\\
\Longrightarrow \left|1 + \frac{\dot{H}}{2H^2}\right|\frac{dS_\mathrm{E}^{(m)}}{dt} =&\, -\frac{n(n-1)}{4}\Omega_nR_\mathrm{h}^n\dot{H}\, ,\nonumber\\
\Longrightarrow \left|1 + \frac{\dot{H}}{2H^2}\right|\frac{dS_\mathrm{E}^{(m)}}{dt} =&\, \frac{n(n-1)}{4}\Omega_n{R_\mathrm{h}}^{n-2}\dot{R}_\mathrm{h}\, ,\nonumber\\
\Longrightarrow \frac{dS_\mathrm{E}^{(m)}}{dt} =&\, -\frac{n(n-1)}{4}\Omega_n\frac{R_\mathrm{h}^n\dot{R}_\mathrm{h}}{\left|1 + \frac{\dot{H}}{2H^2}\right|}\, ,
\label{app-I-1}
\end{align}
where, in the second equality, we use the FLRW Eq.~(\ref{FRW-Einstein}). 
The above equation resembles Eq.~(\ref{E-L-2}). 

We now show that the terms within the square bracket of Eq.~(\ref{MGT-L-3}) cancel with each other. 
Then let us start from
%\begin{align}
%\left[-\frac{d}{dt}\left(\rho_\mathrm{eff}V\right) + \rho_\mathrm{eff}\dot{V} - T\frac{dS_\mathrm{E}^{(m)}}{dt}\right] = -\dot{\rho}_\mathrm{eff}V - %T\frac{dS_\mathrm{E}^{(m)}}{dt}\, ,
%\label{app-II-1}
%\end{align}
%which, due to Eq.~(\ref{AH2}) and the form of $S_\mathrm{E}^{(m)}$ in Eq.~(\ref{E-L-3}), turns out to be,
\begin{eqnarray}
&-&\frac{d}{dt}\left(\rho V + \rho_\mathrm{c}V\right)
+ \left(\rho + \rho_\mathrm{c}\right)\dot{V} - \left(\frac{H}{2\pi}\right)\frac{n(n-1)}{4}\Omega_n {R_\mathrm{h}}^{n-2}\dot{R}_\mathrm{h}\nonumber\\
=&-&\left(\dot{\rho} + \dot{\rho}_\mathrm{c}\right)V - n(n-1)\Omega_n{R_\mathrm{h}}^n\left(\frac{H}{8\pi}\right)\left(\frac{\dot{R}_\mathrm{h}}{{R_\mathrm{h}}^2}\right) \nonumber\\
=&-&n(n-1)\Omega_n{R_\mathrm{h}}^n\dot{H}\left(\frac{H}{8\pi}\right) - n(n-1)\Omega_n
{R_\mathrm{h}}^n\left(\frac{H}{8\pi}\right)\left(\frac{\dot{R}_\mathrm{h}}{{R_\mathrm{h}}^2}\right)\, ,\nonumber\\
=&-&n(n-1)\Omega_n{R_\mathrm{h}}^n\left(\frac{H}{8\pi}\right)\left\{\dot{H} + \frac{\dot{R}_\mathrm{h}}{{R_\mathrm{h}}^2}\right\} = 0\, .
\label{app-II-2}
\end{eqnarray}
The above equation has been used in Sec.~[\ref{sec-law-2-MGT}].

\section*{Acknowledgments}
This work was partially supported by MICINN (Spain), project PID2019-104397GB-I00 and by the program Unidad de Excelencia Maria de Maeztu CEX2020-001058-M, Spain (S.D.O).


\begin{thebibliography}{99}

%\cite{Bekenstein:1973ur}
\bibitem{Bekenstein:1973ur}
J.~D.~Bekenstein,
%``Black holes and entropy,''
Phys. Rev. D \textbf{7} (1973), 2333-2346
doi:10.1103/PhysRevD.7.2333
%5696 citations counted in INSPIRE as of 30 Sep 2022

%\cite{Hawking:1975vcx}
\bibitem{Hawking:1975vcx}
S.~W.~Hawking,
%``Particle Creation by Black Holes,''
Commun. Math. Phys. \textbf{43} (1975), 199-220
[erratum: Commun. Math. Phys. \textbf{46} (1976), 206]
doi:10.1007/BF02345020
%9898 citations counted in INSPIRE as of 30 Sep 2022

%\cite{Bardeen:1973gs}
\bibitem{Bardeen:1973gs}
J.~M.~Bardeen, B.~Carter and S.~W.~Hawking,
%``The Four laws of black hole mechanics,''
Commun. Math. Phys. \textbf{31} (1973), 161-170
doi:10.1007/BF01645742
%2559 citations counted in INSPIRE as of 19 Mar 2022

%\cite{Wald:1999vt}
\bibitem{Wald:1999vt}
R.~M.~Wald,
%``The thermodynamics of black holes,''
Living Rev. Rel. \textbf{4} (2001), 6
doi:10.12942/lrr-2001-6
[arXiv:gr-qc/9912119 [gr-qc]].
%532 citations counted in INSPIRE as of 19 Mar 2022

%\cite{Jacobson:1995ab}
\bibitem{Jacobson:1995ab}
T.~Jacobson,
%``Thermodynamics of spacetime: The Einstein equation of state,''
Phys.\ Rev.\ Lett.\ {\bf 75}, 1260 (1995)
% %doi:10.1103/PhysRevLett.75.1260
[gr-qc/9504004].
%%CITATION = %doi:10.1103/PhysRevLett.75.1260;%%
%1190 citations counted in INSPIRE as of 30 May 2018

%\cite{Padmanabhan:2003gd}
\bibitem{Padmanabhan:2003gd}
T.~Padmanabhan,
%``Gravity and the thermodynamics of horizons,''
Phys.\ Rept.\ {\bf 406}, 49 (2005)
% %doi:10.1016/j.physrep.2004.10.003
[gr-qc/0311036].
%%CITATION = %doi:10.1016/j.physrep.2004.10.003;%%
%396 citations counted in INSPIRE as of 30 May 2018

%\cite{Padmanabhan:2009vy}
\bibitem{Padmanabhan:2009vy}
T.~Padmanabhan,
%``Thermodynamical Aspects of Gravity: New insights,''
Rept.\ Prog.\ Phys.\ {\bf 73}, 046901 (2010)
% %doi:10.1088/0034-4885/73/4/046901
[arXiv:0911.5004 [gr-qc]].
%%CITATION = %doi:10.1088/0034-4885/73/4/046901;%%
%445 citations counted in INSPIRE as of 30 May 2018


%\cite{Hayward:1997jp}
\bibitem{Hayward:1997jp}
S.~A.~Hayward,
%``Unified first law of black hole dynamics and relativistic thermodynamics,''
Class. Quant. Grav. \textbf{15} (1998), 3147-3162
doi:10.1088/0264-9381/15/10/017
[arXiv:gr-qc/9710089 [gr-qc]].
%477 citations counted in INSPIRE as of 06 Sep 2022

%\cite{Cai:2005ra}
\bibitem{Cai:2005ra}
R.~G.~Cai and S.~P.~Kim,
%``First law of thermodynamics and Friedmann equations of
%Friedmann-Robertson-Walker universe,''
JHEP {\bf 0502}, 050 (2005)
[arXiv:hep-th/0501055].


%\cite{Akbar:2006kj}
\bibitem{Akbar:2006kj}
M.~Akbar and R.~G.~Cai,
%``Thermodynamic Behavior of Friedmann Equations at Apparent Horizon
%of FRW Universe,''
Phys.\ Rev.\ D {\bf 75}, 084003 (2007)
% %doi:10.1103/PhysRevD.75.084003
[hep-th/0609128].
%%CITATION = %doi:10.1103/PhysRevD.75.084003;%%
%256 citations counted in INSPIRE as of 24 Apr 2018

%\cite{Cai:2006rs}
\bibitem{Cai:2006rs}
R.~G.~Cai and L.~M.~Cao,
%``Unified first law and thermodynamics of apparent horizon in FRW universe,''
Phys.\ Rev.\ D {\bf 75}, 064008 (2007)
% %doi:10.1103/PhysRevD.75.064008
[gr-qc/0611071].
%%CITATION = %doi:10.1103/PhysRevD.75.064008;%%
%245 citations counted in INSPIRE as of 30 May 2018





%\cite{Akbar:2006er}
\bibitem{Akbar:2006er}
M.~Akbar and R.~G.~Cai,
%``Friedmann equations of FRW universe in scalar-tensor gravity, F(R) gravity and first
%law of thermodynamics,''
Phys.\ Lett.\ B {\bf 635}, 7 (2006)
% %doi:10.1016/j.physletb.2006.02.035
[hep-th/0602156].
%%CITATION = %doi:10.1016/j.physletb.2006.02.035;%%
%216 citations counted in INSPIRE as of 30 May 2018









%\cite{Paranjape:2006ca}
\bibitem{Paranjape:2006ca}
A.~Paranjape, S.~Sarkar and T.~Padmanabhan,
%``Thermodynamic route to field equations in Lancos-Lovelock gravity,''
Phys.\ Rev.\ D {\bf 74}, 104015 (2006)
% %doi:10.1103/PhysRevD.74.104015
[hep-th/0607240].
%%CITATION = %doi:10.1103/PhysRevD.74.104015;%%
%195 citations counted in INSPIRE as of 30 May 2018

%\cite{Sheykhi:2007zp}
\bibitem{Sheykhi:2007zp}
A.~Sheykhi, B.~Wang and R.~G.~Cai,
%``Thermodynamical Properties of Apparent Horizon in Warped DGP Braneworld,''
Nucl.\ Phys.\ B {\bf 779}, 1 (2007)
[arXiv:hep-th/0701198].
%%CITATION = NUPHA,B779,1;%%



%\cite{Jamil:2009eb}
\bibitem{Jamil:2009eb}
M.~Jamil, E.~N.~Saridakis and M.~R.~Setare,
%``Thermodynamics of dark energy interacting with dark matter and radiation,''
Phys.\ Rev.\ D {\bf 81}, 023007 (2010)
% %doi:10.1103/PhysRevD.81.023007
[arXiv:0910.0822 [hep-th]].
%%CITATION = %doi:10.1103/PhysRevD.81.023007;%%
%119 citations counted in INSPIRE as of 30 May 2018

%\cite{Cai:2009ph}
\bibitem{Cai:2009ph}
R.~G.~Cai and N.~Ohta,
%``Horizon Thermodynamics and Gravitational Field Equations in Horava-Lifshitz
%Gravity,''
Phys.\ Rev.\ D {\bf 81}, 084061 (2010)
% %doi:10.1103/PhysRevD.81.084061
[arXiv:0910.2307 [hep-th]].
%%CITATION = %doi:10.1103/PhysRevD.81.084061;%%
%111 citations counted in INSPIRE as of 30 May 2018

%\cite{Wang:2009zv}
\bibitem{Wang:2009zv}
M.~Wang, J.~Jing, C.~Ding and S.~Chen,
%``First law of thermodynamics in IR modified Horava-Lifshitz gravity,''
Phys.\ Rev.\ D {\bf 81}, 083006 (2010)
% %doi:10.1103/PhysRevD.81.083006
[arXiv:0912.4832 [gr-qc]].
%%CITATION = %doi:10.1103/PhysRevD.81.083006;%%
%26 citations counted in INSPIRE as of 30 May 2018

%\cite{Jamil:2010di}
\bibitem{Jamil:2010di}
M.~Jamil, E.~N.~Saridakis and M.~R.~Setare,
%``The generalized second law of thermodynamics in Horava-Lifshitz cosmology,''
JCAP {\bf 1011}, 032 (2010)
% %doi:10.1088/1475-7516/2010/11/032
[arXiv:1003.0876 [hep-th]].
%%CITATION = %doi:10.1088/1475-7516/2010/11/032;%%
%87 citations counted in INSPIRE as of 30 May 2018

%\cite{Gim:2014nba}
\bibitem{Gim:2014nba}
Y.~Gim, W.~Kim and S.~H.~Yi,
%``The first law of thermodynamics in Lifshitz black holes revisited,''
JHEP {\bf 1407}, 002 (2014)
% %doi:10.1007/JHEP07(2014)002
[arXiv:1403.4704 [hep-th]].
%%CITATION = %doi:10.1007/JHEP07(2014)002;%%
%29 citations counted in INSPIRE as of 30 May 2018





%\cite{DAgostino:2019wko}
\bibitem{DAgostino:2019wko}
R.~D'Agostino,
%``Holographic dark energy from nonadditive entropy: cosmological
%perturbations and observational constraints,''
Phys.\ Rev.\ D {\bf 99} (2019) no.10, 103524
%doi:10.1103/PhysRevD.99.103524
[arXiv:1903.03836 [gr-qc]].
%%CITATION = %doi:10.1103/PhysRevD.99.103524;%%
%2 citations counted in INSPIRE as of 14 Jun 2019





%\cite{Sanchez:2022xfh}
\bibitem{Sanchez:2022xfh}
L.~M.~Sanchez and H.~Quevedo,
%``Thermodynamics of the FLRW apparent horizon,''
[arXiv:2208.05729 [gr-qc]].
%0 citations counted in INSPIRE as of 28 Aug 2022


\bibitem{Cognola:2005de}
G.~Cognola, E.~Elizalde, S.~Nojiri, S.~D.~Odintsov and S.~Zerbini,
%``One-loop f(R) gravity in de Sitter universe,''
JCAP \textbf{02} (2005), 010
doi:10.1088/1475-7516/2005/02/010
[arXiv:hep-th/0501096 [hep-th]].

\bibitem{Brevik:2004sd}
I.~H.~Brevik, S.~Nojiri, S.~D.~Odintsov and L.~Vanzo,
%``Entropy and universality of Cardy-Verlinde formula in dark energy universe,''
Phys. Rev. D \textbf{70} (2004), 043520
doi:10.1103/PhysRevD.70.043520
[arXiv:hep-th/0401073 [hep-th]].


%\cite{Nojiri:2023ikl}
\bibitem{Nojiri:2023ikl}
S.~Nojiri and S.~D.~Odintsov,
%``Micro-canonical and canonical description for generalised entropy,''
[arXiv:2304.09014 [gr-qc]].
%0 citations counted in INSPIRE as of 02 Jun 2023


%\cite{Tsallis:1987eu}
\bibitem{Tsallis:1987eu}
C.~Tsallis,
%``Possible Generalization of Boltzmann-Gibbs Statistics,''
J. Statist. Phys. \textbf{52} (1988), 479-487
doi:10.1007/BF01016429
%1405 citations counted in INSPIRE as of 19 Mar 2022

\bibitem{Renyi}
A.~R\'{e}nyi, Proceedings of the Fourth Berkeley Symposium on Mathematics, Statistics and Probability, University of California Press (1960), 547-56.

%\cite{Barrow:2020tzx}
\bibitem{Barrow:2020tzx}
J.~D.~Barrow,
%``The Area of a Rough Black Hole,''
Phys. Lett. B \textbf{808} (2020), 135643
doi:10.1016/j.physletb.2020.135643
[arXiv:2004.09444 [gr-qc]].
%59 citations counted in INSPIRE as of 19 Mar 2022

%\cite{SayahianJahromi:2018irq}
\bibitem{SayahianJahromi:2018irq}
A.~Sayahian Jahromi, S.~A.~Moosavi, H.~Moradpour, J.~P.~Morais Gra\c{c}a, I.~P.~Lobo, I.~G.~Salako and A.~Jawad,
%``Generalized entropy formalism and a new holographic dark energy model,''
Phys. Lett. B \textbf{780} (2018), 21-24
doi:10.1016/j.physletb.2018.02.052
[arXiv:1802.07722 [gr-qc]].
%112 citations counted in INSPIRE as of 19 Mar 2022

%\cite{Kaniadakis:2005zk}
\bibitem{Kaniadakis:2005zk}
G.~Kaniadakis,
%``Statistical mechanics in the context of special relativity. II.,''
Phys. Rev. E \textbf{72} (2005), 036108
doi:10.1103/PhysRevE.72.036108
[arXiv:cond-mat/0507311 [cond-mat]].
%38 citations counted in INSPIRE as of 19 Mar 2022

%\cite{Drepanou:2021jiv}
\bibitem{Drepanou:2021jiv}
N.~Drepanou, A.~Lymperis, E.~N.~Saridakis and K.~Yesmakhanova,
%``Kaniadakis holographic dark energy,''
[arXiv:2109.09181 [gr-qc]].
%11 citations counted in INSPIRE as of 19 Mar 2022

%\cite{Majhi:2017zao}
\bibitem{Majhi:2017zao}
A.~Majhi,
%``Non-extensive Statistical Mechanics and Black Hole Entropy From Quantum Geometry,''
Phys. Lett. B \textbf{775} (2017), 32-36
doi:10.1016/j.physletb.2017.10.043
[arXiv:1703.09355 [gr-qc]].
%56 citations counted in INSPIRE as of 19 Mar 2022

%\cite{Liu:2021dvj}
\bibitem{Liu:2021dvj}
Y.~Liu,
%``Non-extensive Statistical Mechanics and the Thermodynamic Stability of FRW Universe,''
doi:10.1209/0295-5075/ac3f52
[arXiv:2112.15077 [gr-qc]].
%1 citations counted in INSPIRE as of 19 Mar 2022

%\cite{Nojiri:2022aof}
\bibitem{Nojiri:2022aof}
S.~Nojiri, S.~D.~Odintsov and V.~Faraoni,
%``From nonextensive statistics and black hole entropy to the holographic dark universe,''
Phys. Rev. D \textbf{105} (2022) no.4, 044042
doi:10.1103/PhysRevD.105.044042
[arXiv:2201.02424 [gr-qc]].
%4 citations counted in INSPIRE as of 19 Mar 2022

%\cite{Nojiri:2022dkr}
\bibitem{Nojiri:2022dkr}
S.~Nojiri, S.~D.~Odintsov and T.~Paul,
%``Early and late universe holographic cosmology from a new generalized entropy,''
Phys. Lett. B \textbf{831} (2022), 137189
doi:10.1016/j.physletb.2022.137189
[arXiv:2205.08876 [gr-qc]].



%\cite{Odintsov:2023qfj}
\bibitem{Odintsov:2023qfj}
S.~D.~Odintsov and T.~Paul,
%``Generalised (non-singular) entropy functions with applications to cosmology and black holes,''
[arXiv:2301.01013 [gr-qc]].
%1 citations counted in INSPIRE as of 14 Mar 2023


%\cite{Li:2004rb}
\bibitem{Li:2004rb}
M.~Li,
%``A Model of holographic dark energy,''
Phys.\ Lett.\ B {\bf 603} (2004) 1
doi:10.1016/j.physletb.2004.10.014
[hep-th/0403127].
%%CITATION = doi:10.1016/j.physletb.2004.10.014;%%
%1138 citations counted in INSPIRE as of 15 Nov 2019

%\cite{Li:2011sd}
\bibitem{Li:2011sd}
M.~Li, X.~D.~Li, S.~Wang and Y.~Wang,
%``Dark Energy,''
Commun. Theor. Phys. \textbf{56} (2011), 525-604
doi:10.1088/0253-6102/56/3/24
[arXiv:1103.5870 [astro-ph.CO]].
%665 citations counted in INSPIRE as of 10 Mar 2021

%\cite{Wang:2016och}
\bibitem{Wang:2016och}
S.~Wang, Y.~Wang and M.~Li,
%``Holographic Dark Energy,''
Phys.\ Rept.\ {\bf 696} (2017) 1
doi:10.1016/j.physrep.2017.06.003
[arXiv:1612.00345 [astro-ph.CO]].
%%CITATION = doi:10.1016/j.physrep.2017.06.003;%%
%89 citations counted in INSPIRE as of 15 Nov 2019

%\cite{Pavon:2005yx}
\bibitem{Pavon:2005yx}
D.~Pavon and W.~Zimdahl,
%``Holographic dark energy and cosmic coincidence,''
Phys.\ Lett.\ B {\bf 628} (2005) 206
doi:10.1016/j.physletb.2005.08.134
[gr-qc/0505020].
%%CITATION = doi:10.1016/j.physletb.2005.08.134;%%
%436 citations counted in INSPIRE as of 15 Nov 2019

%\cite{Nojiri:2005pu}
\bibitem{Nojiri:2005pu}
S.~Nojiri and S.~D.~Odintsov,
%``Unifying phantom inflation with late-time acceleration: Scalar phantom-non-phantom transition model and generalized holographic dark energy,''
Gen.\ Rel.\ Grav.\ {\bf 38} (2006) 1285
doi:10.1007/s10714-006-0301-6
[hep-th/0506212].
%%CITATION = doi:10.1007/s10714-006-0301-6;%%
%555 citations counted in INSPIRE as of 15 Nov 2019

%\cite{Nojiri:2020wmh}
\bibitem{Nojiri:2020wmh}
S.~Nojiri, S.~D.~Odintsov, V.~K.~Oikonomou and T.~Paul,
%``Unifying Holographic Inflation with Holographic Dark Energy: a Covariant Approach,''
Phys. Rev. D \textbf{102} (2020) no.2, 023540
doi:10.1103/PhysRevD.102.023540
[arXiv:2007.06829 [gr-qc]].
%63 citations counted in INSPIRE as of 14 Jun 2023



%\cite{Gong:2004cb}
\bibitem{Gong:2004cb}
Y.~g.~Gong, B.~Wang and Y.~Z.~Zhang,
%``The Holographic dark energy revisited,''
Phys.\ Rev.\ D {\bf 72} (2005) 043510
doi:10.1103/PhysRevD.72.043510
[hep-th/0412218].
%%CITATION = doi:10.1103/PhysRevD.72.043510;%%
%94 citations counted in INSPIRE as of 15 Nov 2019


%\cite{Khurshudyan:2016gmb}
\bibitem{Khurshudyan:2016gmb}
M.~Khurshudyan,
%``Viscous holographic dark energy universe with Nojiri-Odintsov cut-off,''
Astrophys. Space Sci. \textbf{361} (2016) no.12, 392
doi:10.1007/s10509-016-2981-z
%6 citations counted in INSPIRE as of 09 Mar 2021

%\cite{Nojiri:2022nmu}
\bibitem{Nojiri:2022nmu}
S.~Nojiri, S.~D.~Odintsov and T.~Paul,
%``Modified cosmology from the thermodynamics of apparent horizon,''
Phys. Lett. B \textbf{835} (2022), 137553
doi:10.1016/j.physletb.2022.137553
[arXiv:2211.02822 [gr-qc]].
%7 citations counted in INSPIRE as of 14 Jun 2023

%\cite{Nojiri:2021jxf}
\bibitem{Nojiri:2021jxf}
S.~Nojiri, S.~D.~Odintsov and T.~Paul,
%``Barrow entropic dark energy: A member of generalized holographic dark energy family,''
Phys. Lett. B \textbf{825} (2022), 136844
doi:10.1016/j.physletb.2021.136844
[arXiv:2112.10159 [gr-qc]].
%69 citations counted in INSPIRE as of 14 Jun 2023

%\cite{Landim:2015hqa}
\bibitem{Landim:2015hqa}
R.~C.~G.~Landim,
%``Holographic dark energy from minimal supergravity,''
Int.\ J.\ Mod.\ Phys.\ D {\bf 25} (2016) no.04, 1650050
doi:10.1142/S0218271816500504
[arXiv:1508.07248 [hep-th]].
%%CITATION = doi:10.1142/S0218271816500504;%%
%22 citations counted in INSPIRE as of 16 Nov 2019

%\cite{Gao:2007ep}
\bibitem{Gao:2007ep}
C.~Gao, F.~Wu, X.~Chen and Y.~G.~Shen,
%``A Holographic Dark Energy Model from Ricci Scalar Curvature,''
Phys.\ Rev.\ D {\bf 79} (2009) 043511
doi:10.1103/PhysRevD.79.043511
[arXiv:0712.1394 [astro-ph]].
%%CITATION = doi:10.1103/PhysRevD.79.043511;%%
%334 citations counted in INSPIRE as of 16 Nov 2019

%\cite{Li:2008zq}
\bibitem{Li:2008zq}
M.~Li, C.~Lin and Y.~Wang,
%``Some Issues Concerning Holographic Dark Energy,''
JCAP {\bf 0805} (2008) 023
doi:10.1088/1475-7516/2008/05/023
[arXiv:0801.1407 [astro-ph]].
%%CITATION = doi:10.1088/1475-7516/2008/05/023;%%
%89 citations counted in INSPIRE as of 16 Nov 2019

%\cite{Nojiri:2023nop}
\bibitem{Nojiri:2023nop}
S.~Nojiri, S.~D.~Odintsov and T.~Paul,
%``Holographic realization of constant roll inflation and dark energy: An unified scenario,''
Phys. Lett. B \textbf{841} (2023), 137926
doi:10.1016/j.physletb.2023.137926
[arXiv:2304.09436 [gr-qc]].
%0 citations counted in INSPIRE as of 14 Jun 2023

%\cite{Nojiri:2010wj}
\bibitem{Nojiri:2010wj}
S.~Nojiri and S.~D.~Odintsov,
%``Unified cosmic history in modified gravity: from F(R) theory to Lorentz non-invariant models,''
Phys. Rept. \textbf{505} (2011), 59-144
doi:10.1016/j.physrep.2011.04.001
[arXiv:1011.0544 [gr-qc]].
%3168 citations counted in INSPIRE as of 14 Jun 2023


%\cite{Capozziello:2009nq}
\bibitem{Capozziello:2009nq}
S.~Capozziello, M.~De Laurentis and V.~Faraoni,
%``A Bird's eye view of f(R)-gravity,''
Open Astron. J. \textbf{3} (2010), 49
doi:10.2174/1874381101003020049
[arXiv:0909.4672 [gr-qc]].
%223 citations counted in INSPIRE as of 14 Jun 2023


%\cite{Koivisto:2005yk}
\bibitem{Koivisto:2005yk}
T.~Koivisto,
%``Covariant conservation of energy momentum in modified gravities,''
Class. Quant. Grav. \textbf{23} (2006), 4289-4296
doi:10.1088/0264-9381/23/12/N01
[arXiv:gr-qc/0505128 [gr-qc]].

%\cite{Boehmer:2021aji}
\bibitem{Boehmer:2021aji}
C.~G.~Boehmer and E.~Jensko,
%``Modified gravity: A unified approach,''
Phys. Rev. D \textbf{104} (2021) no.2, 024010
doi:10.1103/PhysRevD.104.024010
[arXiv:2103.15906 [gr-qc]].



\end{thebibliography}
\end{document}